\documentclass[twocolumn]{aastex631}

\usepackage{amsmath}
\usepackage{bm}

\newcommand\baseEDR{25.04}
\newcommand\baseEDRerror{0.02}

\newcommand\baseTRGBTOOL{25.07}
\newcommand\baseTRGBTOOLerror{0.02}

\newcommand\baseTRGBTOOLWU{25.07}
\newcommand\baseTRGBTOOLWUerror{0.01}

\newcommand\baseSeanLi{25.04}
\newcommand\baseSeanLierror{0.01}

\newcommand\fourchipEDR{25.05}
\newcommand\fourchipEDRerror{0.02}

\newcommand\fourchipTRGBTOOL{25.09}
\newcommand\fourchipTRGBTOOLerror{0.01}

\newcommand\fourchipTRGBTOOLWU{25.06}
\newcommand\fourchipTRGBTOOLWUerror{0.01}

\newcommand\fourchipSeanLi{25.05}
\newcommand\fourchipSeanLierror{0.01}

\newcommand\broadcolorEDR{25.05}
\newcommand\broadcolorEDRerror{0.02}

\newcommand\broadcolorTRGBTOOL{25.07}
\newcommand\broadcolorTRGBTOOLerror{0.02}

\newcommand\broadcolorTRGBTOOLWU{25.06}
\newcommand\broadcolorTRGBTOOLWUerror{0.01}

\newcommand\broadcolorSeanLi{25.05}
\newcommand\broadcolorSeanLierror{0.01}

\newcommand\snhostEDR{27.47}
\newcommand\snhostEDRerror{0.05}

\newcommand\mirahostvisitoneEDR{27.17}
\newcommand\mirahostvisitoneEDRerror{0.04}

\newcommand\mirahostvisittwoEDR{27.13}
\newcommand\mirahostvisittwoEDRerror{0.02}

\newcommand\calib{$-$4.362}
\newcommand\calibstat{0.033}
\newcommand\calibsys{0.045}

\begin{document}


\title{Tip of the Red Giant Branch Distances with JWST: An Absolute Calibration in NGC~4258 \\ and First Applications to Type Ia Supernova Hosts}


\author[0000-0002-5259-2314]{Gagandeep S. Anand}
\affiliation{Space Telescope Science Institute, 3700 San Martin Drive, Baltimore, MD 21218, USA}

\author[0000-0002-6124-1196]{Adam G. Riess}
\affiliation{Space Telescope Science Institute, 3700 San Martin Drive, Baltimore, MD 21218, USA}
\affiliation{Department of Physics and Astronomy, Johns Hopkins University, Baltimore, MD 21218, USA}

\author[0000-0001-9420-6525]{Wenlong Yuan}
\affiliation{Department of Physics and Astronomy, Johns Hopkins University, Baltimore, MD 21218, USA}

\author[0000-0002-1691-8217]{Rachael Beaton}
\affiliation{Space Telescope Science Institute, 3700 San Martin Drive, Baltimore, MD 21218, USA}
\affiliation{Department of Physics and Astronomy, Johns Hopkins University, Baltimore, MD 21218, USA}

\author{Stefano Casertano}
\affiliation{Space Telescope Science Institute, 3700 San Martin Drive, Baltimore, MD 21218, USA}

\author[0000-0002-8623-1082]{Siyang Li}
\affiliation{Department of Physics and Astronomy, Johns Hopkins University, Baltimore, MD 21218, USA}

\author[0000-0001-9110-3221]{Dmitry I. Makarov}
\affiliation{Special Astrophysical Observatory of the Russian Academy of Sciences, Nizhnij Arkhyz, Karachay-Cherkessia 369167, Russia}

\author[0000-0003-0736-7609]{Lidia N. Makarova}
\affiliation{Special Astrophysical Observatory of the Russian Academy of Sciences, Nizhnij Arkhyz, Karachay-Cherkessia 369167, Russia}

\author[0000-0002-9291-1981]{R. Brent Tully}
\affiliation{Institute for Astronomy, University of Hawaii, 2680 Woodlawn Drive, Honolulu, HI 96822, USA}

\author[0000-0001-8089-4419]{Richard I. Anderson}
\affiliation{Institute of Physics, \'Ecole Polytechnique F\'ed\'erale de Lausanne (EPFL),\\ Observatoire de Sauverny, 1290 Versoix, Switzerland}

\author[0000-0003-3889-7709]{Louise Breuval}
\affiliation{Department of Physics and Astronomy, Johns Hopkins University, Baltimore, MD 21218, USA}

\author[0000-0001-8416-4093]{Andrew Dolphin}
\affiliation{Raytheon, 1151 E. Hermans Road, Tucson, AZ 85756}
\affiliation{University of Arizona, Steward Observatory, 933 North Cherry Avenue, Tucson, AZ 85721, USA}

\author{Igor D. Karachentsev}
\affiliation{Special Astrophysical Observatory of the Russian Academy of Sciences, Nizhnij Arkhyz, Karachay-Cherkessia 369167, Russia}

\author[0000-0002-1775-4859]{Lucas M. Macri}
\affiliation{NSF's NOIRLab, 950 N Cherry Ave, Tucson, AZ 85719, USA}

\author[0000-0002-4934-5849]{Daniel Scolnic}
\affiliation{Department of Physics, Duke University, Durham, NC 27708, USA}


\begin{abstract}
The tip of the red giant branch (TRGB) allows for the measurement of precise and accurate distances to nearby galaxies, based on the brightest ascent of low-mass red giant branch stars before they undergo the helium flash. With the advent of JWST, there is great promise to utilize the technique to measure galaxy distances out to at least 50~Mpc, significantly further than HST's reach of 20~Mpc. However, with any standard candle, it is first necessary to provide an absolute reference. Here we use Cycle 1 data to provide an absolute calibration in the F090W filter. F090W is most similar to the F814W filter commonly used for TRGB measurements with HST, which had been adopted by the community due to minimal dependence from the underlying metallicities and ages of stars. The imaging we use was taken in the outskirts of NGC~4258, which has a direct geometrical distance measurement from the Keplerian motion of its water megamaser. Utilizing several measurement techniques, we find $M_{TRGB}^{F090W}$ = \calib~$\pm$ \calibstat~(stat) $\pm$ \calibsys~(sys)~mag (Vega) for the metal-poor TRGB. We also perform measurements of the TRGB in two Type Ia supernova hosts, NGC~1559, and NGC~5584. We find good agreement between our TRGB distances and previous distance determinations to these galaxies from Cepheids ($\Delta$ = 0.01 $\pm$ 0.06~mag), with these differences being too small to explain the Hubble tension ($\sim$0.17~mag). As a final bonus, we showcase the serendipitous discovery of a faint dwarf galaxy near NGC~5584.
\end{abstract}




\section{Introduction} \label{sec:intro}

The tip of the red giant branch (TRGB) has become an increasingly prolific way to measure precise and accurate distances to nearby galaxies \citep{1993ApJ...417..553L, 2006AJ....131.1361K, 2007ApJ...661..815R, 2015ApJ...807..133J, 2017AJ....154...51M, 2019ApJ...885..141B, 2020ApJ...898...57D, 2021ApJ...914L..12S, 2021MNRAS.501.3621A}. With the Advanced Camera for Surveys (ACS) instrument presently operating on the Hubble Space Telescope (HST), measurements of galaxy distances accurate to 5$\%$ can be performed out to 10~Mpc with a single orbit of time. Indeed, TRGB distances to $\sim$500 galaxies have been obtained and uniformly analyzed under the umbrella of the Extragalactic Distance Database's \citep{2009AJ....138..323T} Color-Magnitude Diagrams and Tip of the Red Giant Branch Distances Catalog\footnote{Available for public access at \url{edd.ifa.hawaii.edu}} \citep{2009AJ....138..332J, 2021AJ....162...80A}. 

\begin{deluxetable*}{ccccc}[t]
\tabletypesize{\small}
\tablewidth{0pt}
\tablehead{
\colhead{Date} & \colhead{Epoch} & \colhead{Exposure} & \colhead{SW Filter} & \colhead{Exp. time [s]}}
\startdata
2023-01-30 & N5584-- Epoch 1 &  009001\_03101\_* &  F090W &   418.7$\times4$ \\
2023-01-30 & N5584-- Epoch 1 &  009001\_05101\_* &  F150W &   526.1$\times4$ \\
2023-02-21 & N5584-- Epoch 2 &  010001\_02101\_* &  F090W &   418.7$\times4$ \\
2023-02-21 & N5584-- Epoch 2 &  010001\_02103\_* &  F150W &   526.1$\times4$ \\
2023-05-02 & N4258-- Epoch 1 &  005001\_03101\_* &  F090W &   257.7$\times4$ \\
2023-05-02 & N4258-- Epoch 1 &  005001\_03103\_* &  F150W &   365.1$\times4$ \\
2023-05-17 & N4258-- Epoch 2 &  006001\_03101\_* &  F090W &   257.7$\times4$ \\
2023-05-17 & N4258-- Epoch 2 &  006001\_03103\_* &  F150W &   365.1$\times4$ \\
2023-06-30 & N1559-- Epoch 1 &  001001\_02101\_* &  F090W &   418.7$\times4$ \\
2023-06-30 & N1559-- Epoch 1 &  001001\_04101\_* &  F150W &   526.1$\times4$ \\
2023-07-15 & N1559-- Epoch 2 &  002001\_03101\_* &  F090W &   418.7$\times4$ \\
2023-07-15 & N1559-- Epoch 2 &  002001\_03103\_* &  F150W &   526.1$\times4$ \\
\enddata
\caption{Observation log for the data used in this paper. All data was re-run with the \texttt{jwst\_1130.pmap} or greater context version. The name of all observations start with ``jw01685".  \label{tb:obs}}
\end{deluxetable*}

The underlying premise for its use as a standard candle is relatively straightforward$-$ stars continue to fuse hydrogen into helium within a shell around their inert helium cores as they ascend the red giant branch. For low-mass stars ($0.5 < M < 2 M_{\odot})$, once the helium core reaches a critical mass and temperature, the conditions become suitable for ignition of helium fusion within the core, and the star rapidly rearranges its structure and becomes a much fainter horizontal branch star. Importantly, the luminosity at the TRGB is essentially independent of the initial stellar mass of the low-mass precursor, allowing us to exploit it as a standard candle. In practice, complications come from 1) the effects of line-blanketing, 2) varying populations of AGB stars right near and above the TRGB, and 3) the presence of dust and crowding within the host galaxy, among others. These issues can be greatly reduced, if not eliminated, by 1) performing measurements in filters where the magnitude of the TRGB has little-to-no variation with underlying metallicity over a broad range of observed colors, and 2+3) performing measurements in the uncrowded outer regions of galaxies (or alternatively, carefully correcting for crowding via artificial star experiments). More detailed discussions of the use of the TRGB as a standard candle can be found in the literature (e.g. \citealt{2017A&A...606A..33S, 2018SSRv..214..113B}).

With the advent of JWST \citep{Rigby2023, Gardner2023} and its NIRCam imager \citep{2023PASP..135b8001R}, there is great promise to extend the reach of the TRGB out to much further distances due to its greater sensitivity, sharper resolution, and the increased intrinsic brightness of red giants in the near-infrared. Indeed, a look at the Early Release Science observations of WLM \citep{Weisz2023} shows that JWST/NIRCam is capable of reaching down to at least $m_{F090W}$ = 30 mag, allowing TRGB measurements out to at least 50~Mpc (while allowing for photometry that reaches one magnitude below the TRGB). 

As we will show, even relatively modest exposure times can allow for substantially further measurements of the TRGB than can be done with equivalent exposure times on HST (even after accounting for the increased JWST overheads). To allow for measurements of galaxy distances via the TRGB with JWST observations, we present an initial \textit{absolute} calibration of the F090W TRGB magnitude with Cycle 1 data taken in the megamaser host galaxy NGC~4258. We also highlight the effectiveness of F090W for JWST TRGB measurements by showcasing applications to two type Ia supernova host galaxies (NGC~1559 and NGC~5584) which lie at $\sim$20~Mpc.


\section{Data} \label{sec:data}

\begin{figure}
\epsscale{1.1}
\plotone{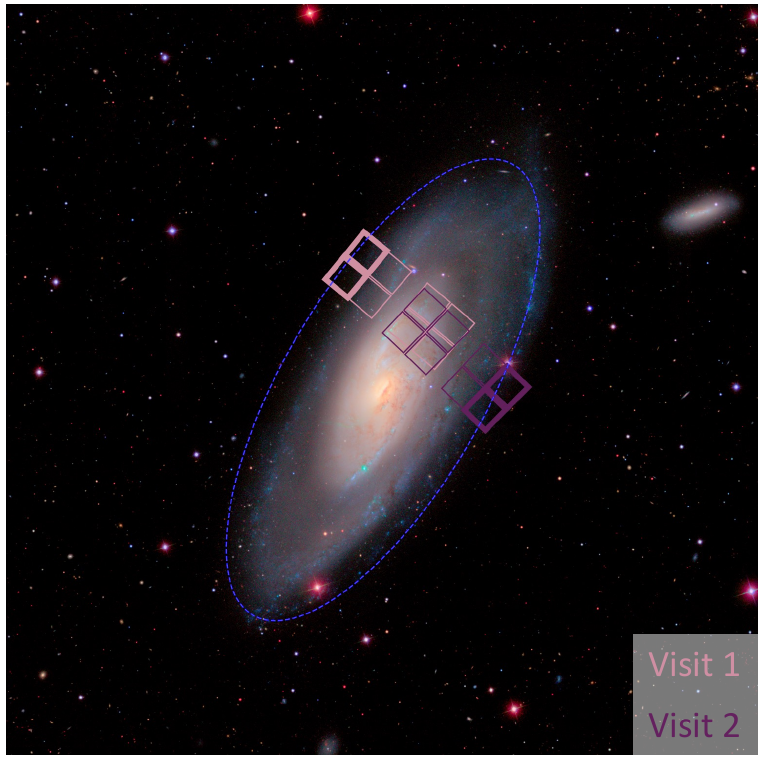}
\caption{An SDSS \citep{SDSS2015} \textit{gri} color composite of NGC~4258, where North is up and East is to the left. Overlain on the SDSS imaging are the footprints of the two visits of our observations in NGC~4258 covered by JWST GO-1685. The dashed blue line indicates $D_{25}$, or the 25th B-band magnitude isophote from \cite{1991rc3..book.....D}-- only stars external to this radius are used for our baseline TRGB measurements. All stars found within the outermost chips from each visit (in bold) were used for our spatial selection variant (see Table \ref{tb:trgb}).}
\label{footprint}
\end{figure}

The data used in this paper was obtained via the Cycle 1 JWST proposal GO-1685 \citep{1685prop}. The main objective of this program is to obtain precise photometry of Cepheid variables, Mira variables, and red giant branch stars, with the goal of reducing systematics along the distance ladder, and as a result, in measurements of the Hubble Constant. One of the galaxies targeted with this program was NGC~4258, which is host to a water megamaser which allows for the determination of a highly precise geometric distance to this galaxy \citep{2013ApJ...775...13H, 2019ApJ...886L..27R}. The data was taken in two separate visits, with the F090W, F150W, and F277W filters, although we do not use the long-wavelength (F277W) images in this paper (see the discussion in \S \ref{sec:data-redux}). We have re-processed all the underlying data to the \texttt{jwst\_1130.pmap} or greater context files\footnote{See full descriptions of updates at \url{jwst-crds.stsci.edu}.}, which include recent updates to the NIRCam flat-fields introduced in \texttt{jwst\_1125.pmap}, as well as to the zeropoints introduced in \texttt{jwst\_1126.pmap}. We note that our photometry is performed using the Vega-Vega zeropoints, and not the Sirius-Vega zeropoints\footnote{See the explanations provided at \url{https://jwst-docs.stsci.edu/jwst-near-infrared-camera/nircam-performance/nircam-absolute-flux-calibration-and-zeropoints}.}. Notably, the uncertainties in the flux calibrations with this latest suite of updates are \textit{``now less than 1 percent for most filter/detector combinations"}, including those used with our program. Table \ref{tb:obs} contains additional details about the individual observations, and Figure \ref{footprint} shows footprints of our NGC~4258 JWST observations overlaid on SDSS \textit{gri} imaging \citep{SDSS2015}.

\begin{figure*}[t!]
\epsscale{1.1}
\plottwo{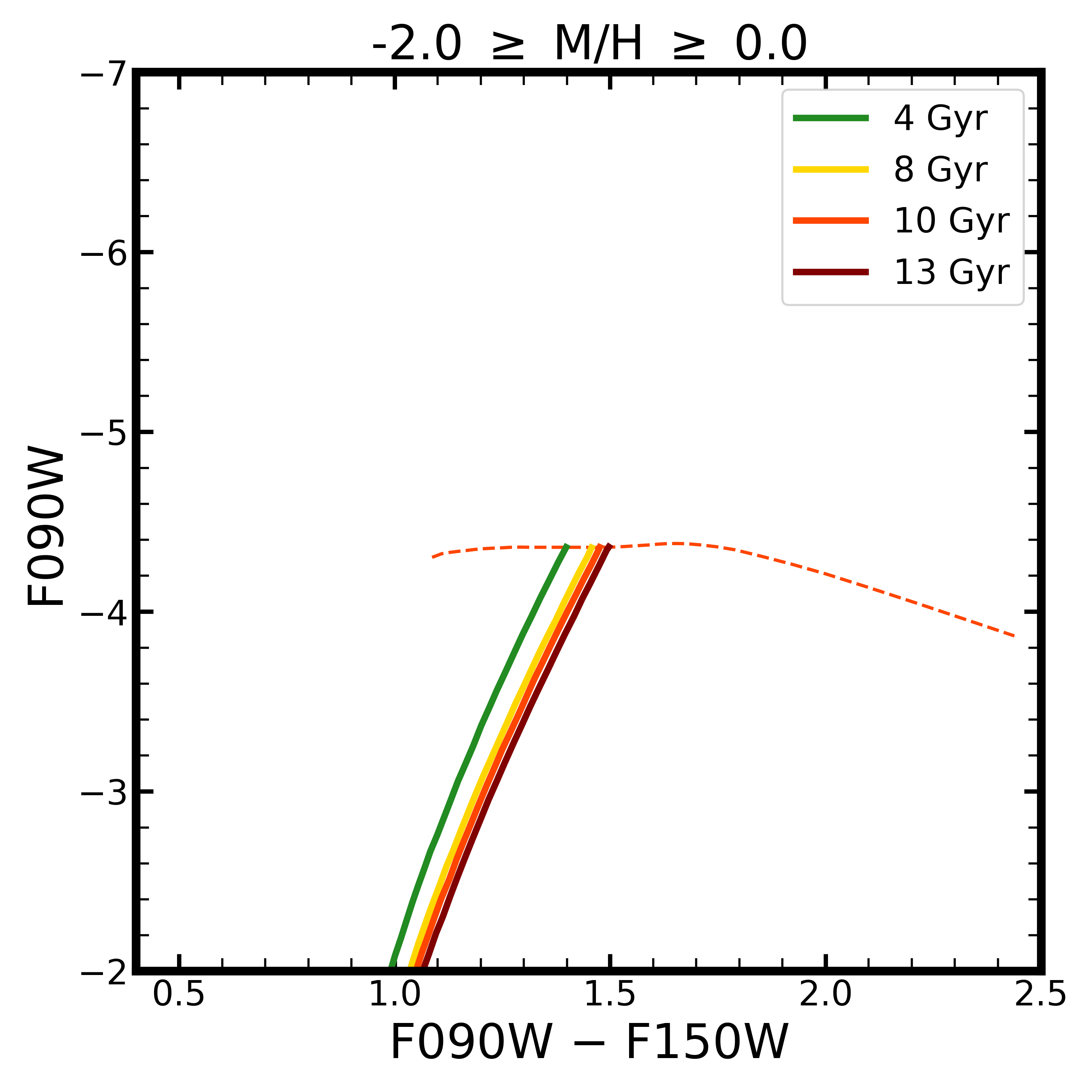}{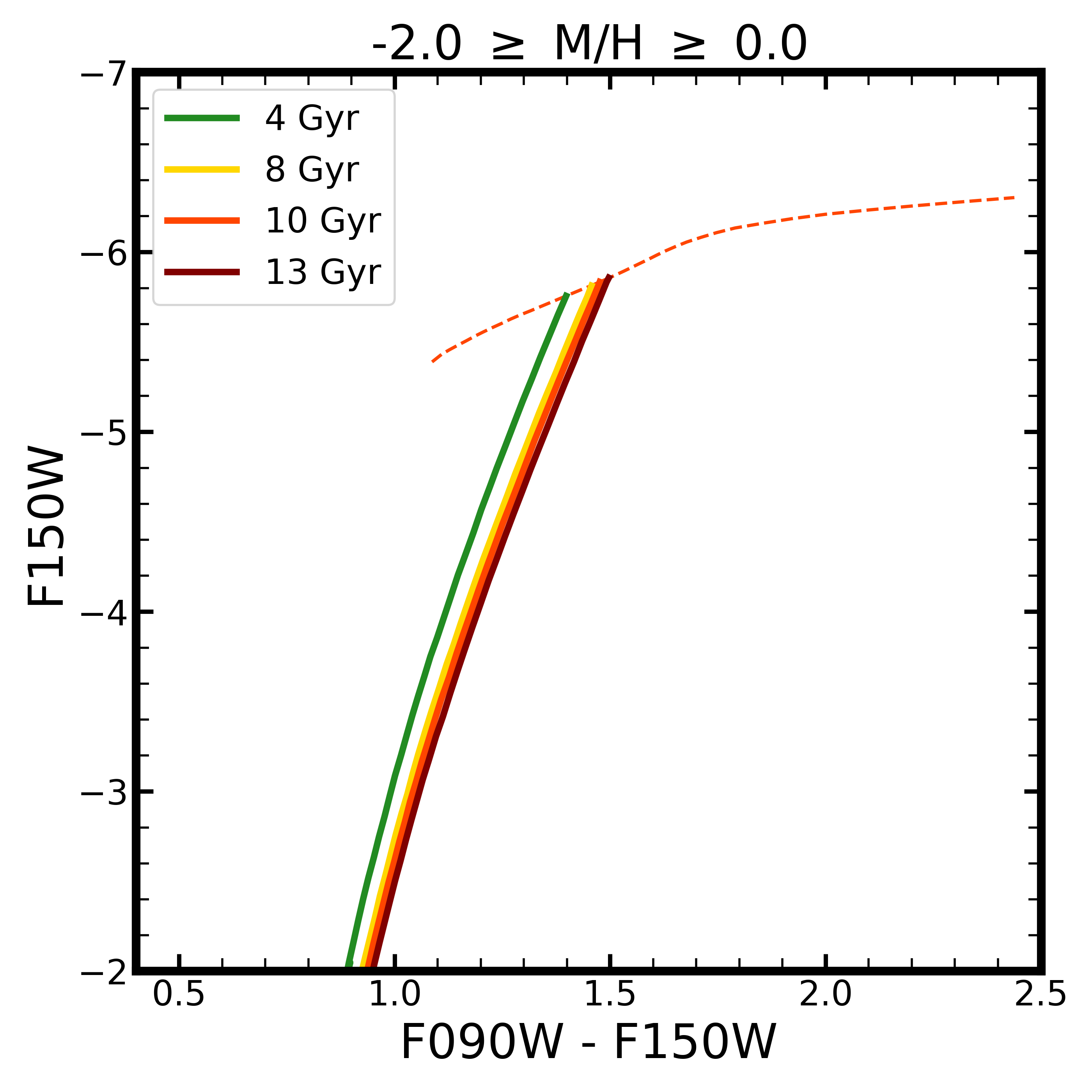}
\caption{PARSEC red giant branch isochrones in our choice of JWST filters \citep{2012MNRAS.427..127B}. The solid lines show populations of varying underlying ages, and the dashed orange line shows the predicted variation of the absolute magnitude of the TRGB as a function of metallicity ($-$2.0 $\ge$ M $\ge$ 0.0 dex) for the 10 Gyr population. In F090W, there is minuscule variation with age over the range of 4$-$13~Gyr ($<$ 0.005 mag), and still little variation with metallicity ($<$ 0.02 mag) over the color range of F090W$-$F150W = 1.15$-$1.75 mag. The situation in F150W is much different, where the absolute magnitude of the TRGB is a sharp function of age and metallicity.}
\label{parsec}
\end{figure*}

The inclusion of F090W imaging in our program allows us to measure the TRGB in a filter where there is little-to-no expected variation of the TRGB over a modest range of colors (the metallicity and age variations are projected onto the color). This is very similar to the case with HST's F814W filter (see Figure 1 in \citealt{2020ApJ...891...57F}), which has been by far the most popular filter with which to perform HST TRGB measurements \citep{2009AJ....138..332J, 2021AJ....162...80A}. The inclusion of the secondary filter F150W allows us to generate a color-magnitude diagram (CMD) and prevent contamination with young stellar populations (e.g. main-sequence stars and supergiants). Figure \ref{parsec} shows PARSEC \citep{2012MNRAS.427..127B} red giant branch isochrones in our choice of JWST filters for a range of stellar ages (solid lines). It can be appreciated that the anticipated brightness of the TRGB in F090W has effectively zero ($<$ 0.005 mag) dependence on stellar age for sufficiently old ($\ge$ 4 Gyr) stellar populations, which to our knowledge of galaxy formation and evolution populate essentially every nearby galaxy\footnote{There are less than a handful of \textit{candidate} young galaxies in the local Universe, including the Peekaboo dwarf galaxy \citep{2023MNRAS.518.5893K}.}. Also shown is the dependence of the F090W magnitude of the TRGB on the underlying stellar metallicity (for a 10 Gyr stellar population, in the dashed line). There is great constancy up until F090W-F150W $\sim$1.8 mag, after which the TRGB begins to tilt downwards. In particular, between F090W-F150W = 1.15 $-$ 1.75, there is only $\sim$0.02 mag of variation in the absolute magnitude of the TRGB in F090W. The same plot is repeated on the right-hand side of Figure \ref{parsec}, except with F150W as the y-axis. It can be seen that while the absolute magnitude of the TRGB is substantially brighter, there is also a large variation with both metallicity and age. While the precise details (such as the predicted absolute magnitude) differ amongst different families of isochrones (e.g. BaSTI; \citealt{2004ApJ...612..168P}, MIST; \citealt{2016ApJ...823..102C}), the overall picture is the same -- F090W is an incredibly stable place to measure the TRGB.

\begin{figure*}[]
\epsscale{1.1}
\plotone{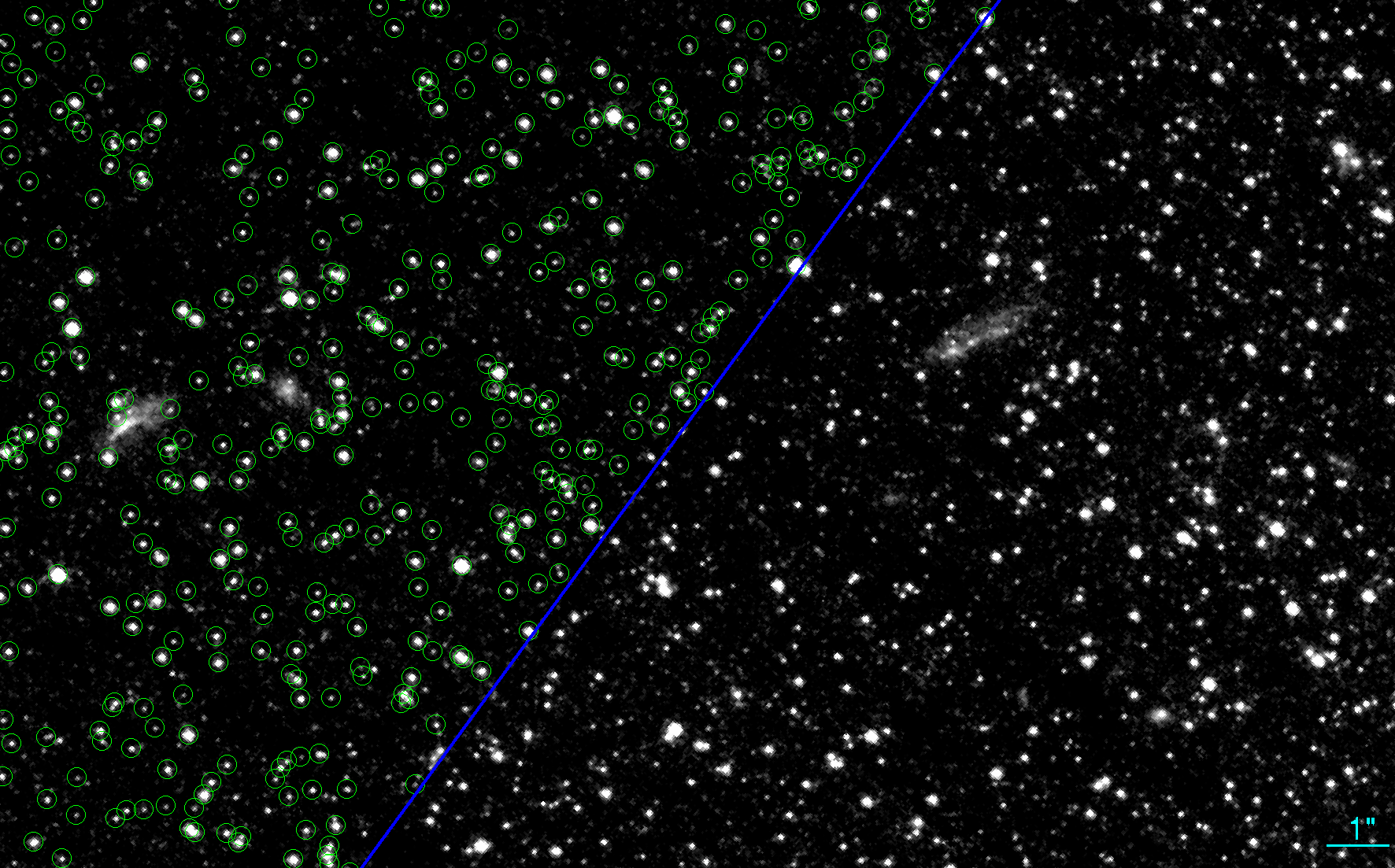}
\caption{A 22$\arcsec\times$14$\arcsec$ cutout from our F150W reference frame image of Visit 1 shown with an arcsinch stretch. The green circles represent stellar detections which pass our quality criteria and a radial selection exterior to $D_{25}$ (shown in blue). A 1$\arcsec$ cyan bar is shown for scale in the bottom right. It can be appreciated that stars in the final list are well-resolved and are located in a relatively uncrowded portion of the outer regions of NGC~4258.}
\label{cutout}
\end{figure*}

Our chosen filter-set is in contrast to another Cycle 1 program \citep{2021jwst.prop.1995F} that aims in part to measure TRGB distances to nearby SN Ia host galaxies with F115W as the primary filter. With F115W, there is substantial variation over even small ranges of colors in the observed magnitude of the TRGB (about half as much as seen with F150W in Figure \ref{parsec}). While this slope may be ``rectified" before making the measurement \citep{2009ApJ...690..389M, 2018ApJ...858...11M}, such a process increases the uncertainty in the underlying measurements. And though there is some benefit due to the increased brightness of the TRGB in F115W when compared with F090W, the main benefit of reduced exposure times may not be worth the increased uncertainties, especially in cases where the highest fidelity TRGB measurements are desired (e.g. for the eventual purposes of measuring the Hubble Constant). Additionally, for more nearby galaxies (within $\sim$20 Mpc), JWST visit times are largely overhead dominated, and filter choice makes a relatively small difference to the total charged times. Indeed, a Cycle 2 program \citep{2023jwst.prop.3055T} which aims to connect the TRGB and SBF distance scales (as an alternative to the traditional Cepheid and SNe Ia route) is set to use the same filter set as the data from GO-1685. We note that the JWST Early Release Science (ERS) Program for Resolved Stellar Populations \citep{Weisz2023} utilizes both of our chosen filters, indicating there is strong community support for their use for general resolved stellar populations science. 


\section{Data Reduction} \label{sec:data-redux}

We perform PSF photometry on the underlying stage 2 images (\texttt{*cal.fits}), with the stage 3 F150W image (\texttt{*i2d.fits}) as the reference frame used for source detection and mutual image alignment. We treat the photometry for each visit independently, except for NGC~5584 (which has overlapping visits in its halo, and greatly benefits from the doubled exposure time). Our choice of software for PSF photometry is DOLPHOT \citep{2000PASP..112.1383D,2016ascl.soft08013D}, along with its NIRCam module \citep{Weisz2023, Weiszinprep}. Specifically, we use the latest major release version of DOLPHOT and its NIRCam PSFs, which are from April 6th, 2023. We note that there were two minor updates since then. The first was on October 11, 2023, and this update only changed the default zeropoints to the ``Sirius-Vega" system. However, we retain the usage of the original ``Vega-Vega" magnitude system. A second update was provided on December 2, 2023, which provided the ``$-$etctime" option for adjusting the exposure times in the image headers to be in line with the ETC values. As we will discuss shortly, changes brought upon by use of this option are not particularly relevant for the very high S/N data to be presented for NGC~4258.

We follow the recommendations for reduction parameters outlined in the DOLPHOT NIRCam manual\footnote{http://americano.dolphinsim.com/dolphot/dolphotNIRCam.pdf}. We note that the pixel scale of the long-wavelength channel (0.063$\arcsec$) is much coarser than that of the short-wavelength channel (0.031$\arcsec$), resulting in data that is of noticeably worse quality. Combined with fact that there are currently minor but non-zero photometric impacts by running simultaneous short-wavelength and long-wavelength photometry\footnote{At present, these can be greatly reduced, but not entirely eliminated with the use of the ``warmstart" option-- see \cite{2023ApJ...956L..18R} for a lengthier discussion.}, we opt to analyze only the short-wavelength NIRCam data in this work. 

A great deal of the initial output photometry from DOLPHOT contains detections of low quality. To cull our initial photometry and generate a high-quality source list, we adopt a modified version of the quality cuts given by \cite{Warfield2023}, which were developed using the ERS data from \cite{Weisz2023}. Specifically, we select for sources with: (1) Crowding $<$ 0.5; (2) $\mathrm{Sharpness^{2}}$ $\le$ 0.01; (3) Object Type $\le$ 2; (4) S/N $\ge$ 5; (5) Error Flag $\le$ 2. The quality cuts are applied to both filters (F090W and F150W), except for object type, which is not filter-specific. An image cutout from the first visit with the stellar detections overlaid can be seen in Figure \ref{cutout} (after applying the spatial cut described in \S 4.1). We provide our DOLPHOT photometry on Github\footnote{\url{https://github.com/gsanand/anand24_jwst_trgb}}.

To quantify the levels of completeness, photometric error, and bias present in our data, we perform artificial star experiments within DOLPHOT. One at a time, we insert and recover $\sim$100,000 stars of varying ranges of input magnitudes and colors for each chip and record these results. The artificial stars are processed with the same quality cuts as applied to the genuine stellar photometry. 

\begin{figure}
\epsscale{1.1}
\plotone{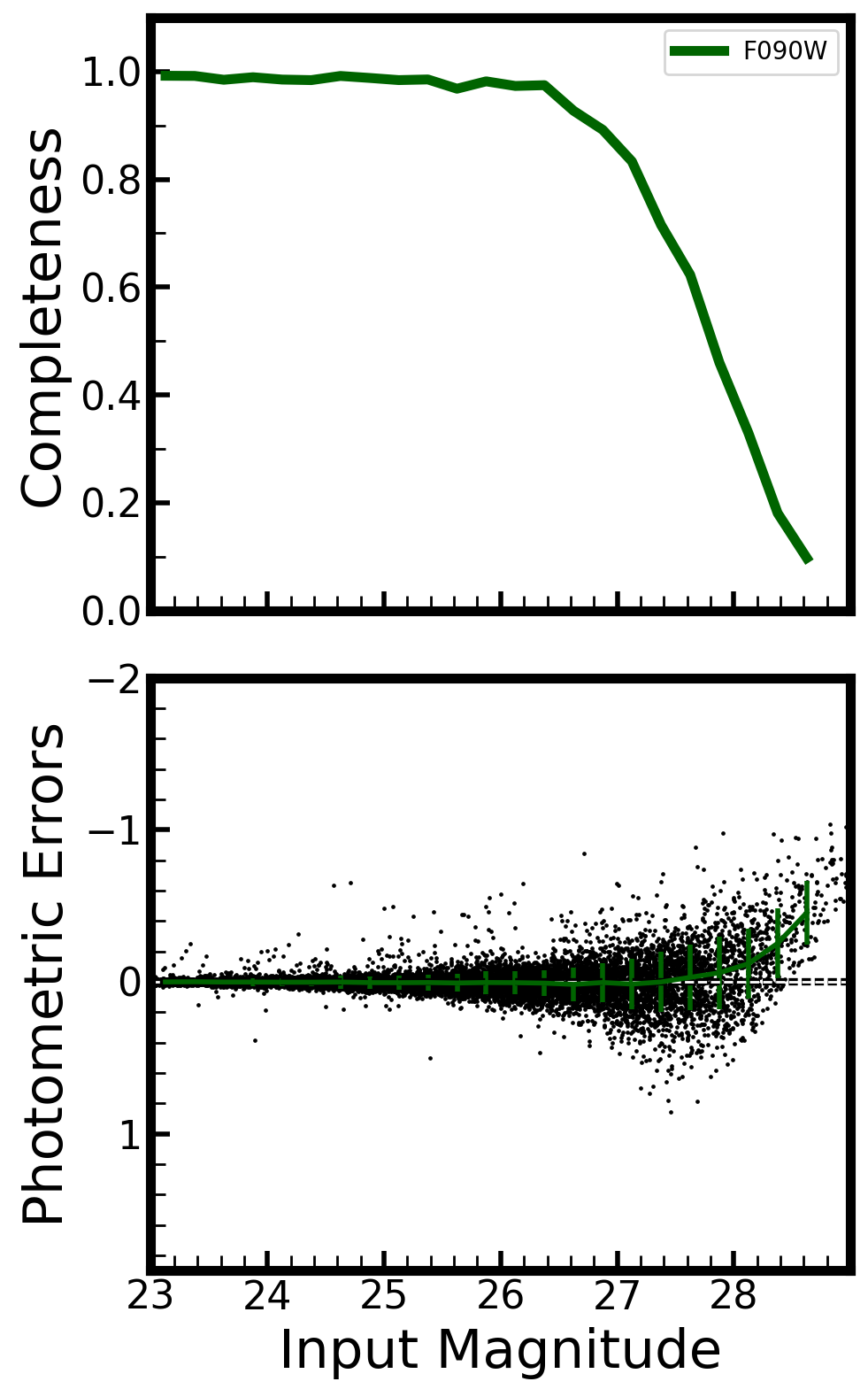}
\caption{Results from F090W artificial star experiments for NGC~4258 used for our baseline TRGB results (though only 2 out of 4 methods utilize the artificial stars). The top panel shows the completeness curve, which measures the percentage of injected stars that were successfully recovered (including meeting all of our quality criteria). The bottom panel shows the offset between injected and recovered stars (black points), as well as the overall photometric bias (green line). The key point here is that the effects of photometric completeness and bias are entirely negligible near the magnitude of the TRGB ($m_{F090W}$ $\sim$ 25.0 mag). Note that the TRGB measurement tools described later do not use the simple binned versions of these curves shown here, but instead smoothed versions to avoid errors from their apparent jaggedness.}
\label{curves}
\end{figure}

An important note, of which we became aware of towards the end of writing this manuscript, is that there are discrepancies between the JWST exposure times provided within the image headers when compared with those provided by the JWST Exposure Time Calculator (the former can be over ten percent longer in some instances). This mismatch leads to some ambiguity in the determination of the true photometric uncertainties determined via both Poisson noise characteristics and artificial star experiments, especially in parts of the CMD with low S/N. However, we note that the NGC~4258 data is of exceptionally high S/N near the magnitude of the TRGB, where these concerns should have a negligible overall impact. As shown in Figure \ref{curves}, the photometric completeness curve remains near 100$\%$ over one magnitude \textit{below} the TRGB. Similarly, photometric bias is not a concern until we reach much fainter magnitudes. The same may or may not be said for our supernova host galaxies, for which the TRGB appears at much fainter magnitudes (see Section 6 for more details). For this reason, we only present edge-detection measurements of the TRGB for our supernova hosts, and reserve a more detailed analysis including methods which rely on artificial star results to a later paper.

\begin{figure*}
\epsscale{1.1}
\plotone{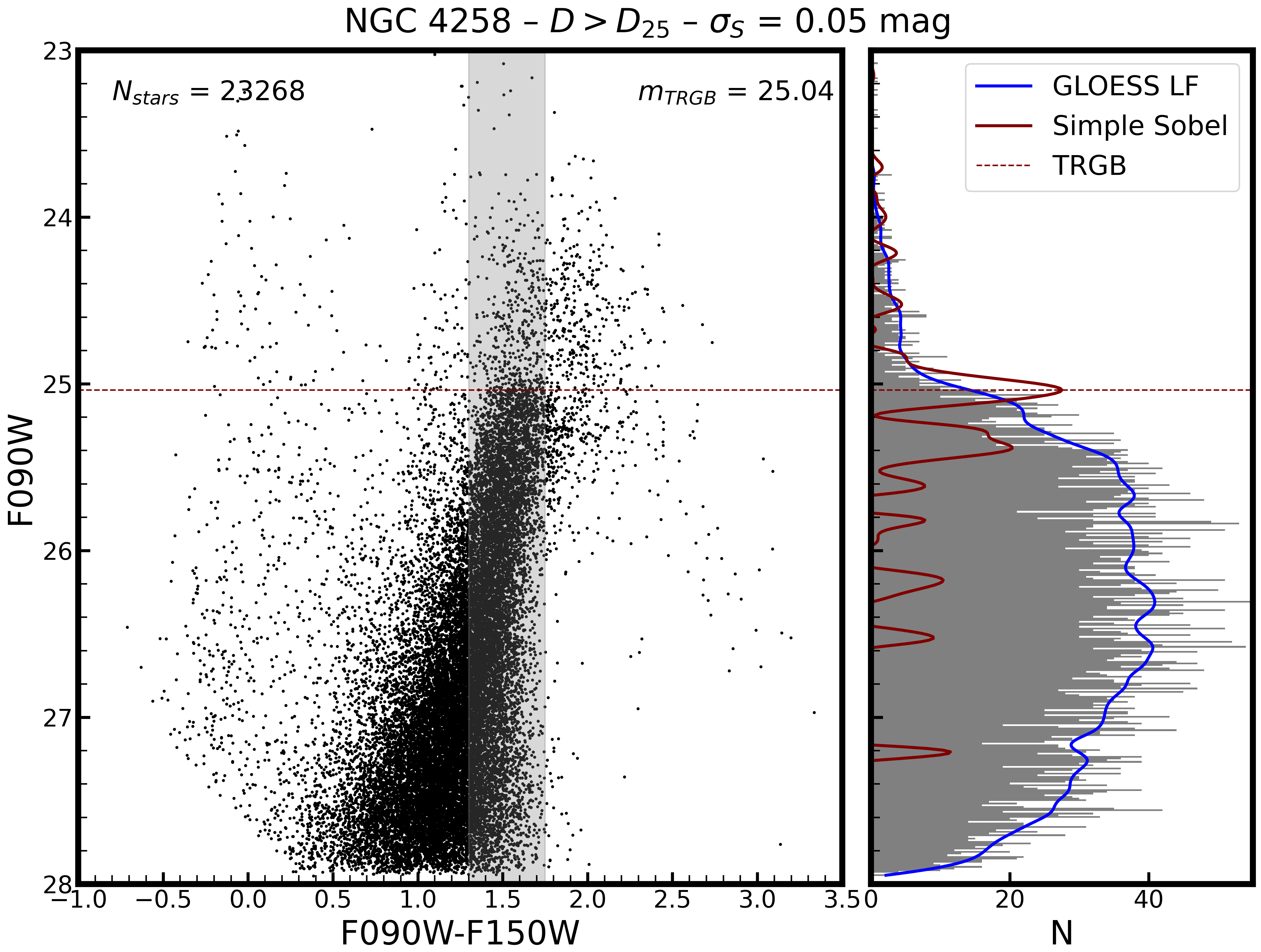}
\caption{Our baseline edge detection result for the TRGB measurement in NGC~4258. The left panel shows the underlying color magnitude diagram from our spatially trimmed region, and the grey band shows our limited color selection region. In the right hand panel, the binned (grey) and GLOESS smoothed (blue) luminosity functions are visible, as well as the output from the Sobel filter (maroon). The data near the TRGB is of exceptionally high quality, as it is nearly 3 magnitudes brighter than the faint limit of S/N = 5.}
\label{edge-detect}
\end{figure*}


\section{Analysis} \label{sec:analysis}

Broadly speaking, there are two distinct methods that have been used in the literature to perform a measurement of the TRGB. The first set involves using an edge-detection algorithm (typically a Sobel filter) to locate the discontinuity that is to be expected with the onset of the TRGB \citep{1993ApJ...417..553L,1999ApJ...511..671S, 2008ApJ...689..721M, 2018ApJ...852...60J}. This is a conceptually simple technique, though there are differences between applications such as the bin-width used to construct the luminosity function and whether or not the luminosity function is smoothed before applying an edge detector (and if so, how). The second methodology involves fitting a model luminosity function to the data. This approach was introduced by \cite{mendez2002}, where they adopt a theoretical luminosity function of the general form\footnote{The definitions of a, b, and c differ between \cite{mendez2002} and \cite{2006AJ....132.2729M}, and we choose to adopt the definitions from the latter.}
\begin{equation}
\psi = \left\{\begin{matrix}10^{a(m-m_{TRGB})+b}, m-m_{TRGB} \geq 0 \\ 
\\
10^{c(m-m_{TRGB})}, \ \ \ m-m_{TRGB} < 0 \end{matrix}\right. 
\end{equation}
where a is the power-law slope for stars below the TRGB, c is the power-law slope for stars above the TRGB, and b is the strength of the discontinuity. Further enhancements presented in \cite{2006AJ....132.2729M} introduce the ability to incorporate results from artificial star experiments, which allows one to account for photometric bias, completeness, and errors present in the data. \cite{2014AJ....148....7W} present a modified version using a non-linear least squares method based on a Levenberg-Marquardt algorithm, instead of the Broyden, Fletcher, Goldfarb, Shanno (BFGS) maximum-likelihood algorithm used by \cite{2006AJ....132.2729M}. 

In this work, we will compare results from several iterations of both of the above methods while using the same underlying photometry, which will highlight any differences introduced by adopting distinct measurement techniques. 

\subsection{Spatial Selections}

A look at Figure \ref{footprint} shows that the majority of our JWST observations lie within the disk of NGC~4258 (intentionally, as to measure Cepheid variables). Many previous works have discussed the importance of limiting TRGB measurements to the regions of galaxies which are relatively uncrowded and unaffected by dust internal to the host galaxy \citep{2006AJ....132.2729M,2018AJ....156..105A, 2019ApJ...885..141B,2021ApJ...915...34H}. For NGC~4258 specifically, \cite{2021ApJ...906..125J} adopt a radial selection of stars outside a semi-major axis cut of 14$\arcsec$, in part guided by a relatively high-sensitivity HI map from \cite{2011A&A...526A.118H}. However, we note that such high-sensitivity HI data is not available for many nearby galaxies (including many SN Ia hosts), and thus presents difficulties when trying to adopt similar spatial cuts for TRGB measurements in other galaxies. 

In this work, we adopt the same, simple radial selection as \cite{2022ApJ...932...15A}, namely the 25th B-magnitude isophote from \cite{1991rc3..book.....D} ($D_{25}$) as shown by the dashed-blue line in Figure \ref{footprint}. We find 23,268 sources which pass our quality cuts outside this region (for reference, there are $\sim$100,000 sources within an individual NIRCam SW chip for observations centered on the disk of NGC~4258). We note that isophotal radii are more simple to measure from existing all-sky ground-based surveys, and can provide a more uniform method of spatial selections (as they account for differing sizes of galaxies). 

Another potential avenue of performing spatial cuts relies on selection regions with relatively low numbers of young, main-sequence stars in their CMDs \citep{2021AJ....162...80A, 2023ApJ...954...87W, 2023arXiv230610103L, 2023ApJ...954L..31S}. We note that the radial cut we adopt for this work results in a relatively low ratio of main-sequence/giant branch stars. Specifically, blue stars (crudely defined as those with F090W-F150W $<$ 0.5 mag, based on the appearance of the main sequence stars in the CMD) make up less than 1 out of every 28 stars within a magnitude range of 0.5 mag above and below the TRGB, indicating there is minimal contamination from young stellar populations. Regardless, to further reduce the level of contamination from young stellar populations, we will adopt a color selection to the CMD before performing our TRGB measurements. Additionally, we will also present results based on just the outermost 2 of 8 chips from each visit (effectively a crude spatial cut) to test our sensitivity to the adopted spatial criterion.

Lastly, we make sure to perform the same set of spatial selections on the artificial stars, as to not create a mismatch between the environments in which the selected genuine stars and artificial stars reside.

\subsection{Color Selections}

Earlier we mention that there is only $\sim$0.02 mag of anticipated variation in the absolute magnitude of the TRGB in F090W between colors of F090W-F150W = 1.15$-$1.75 mag. This would provide a natural color selection for which to limit our TRGB analysis with. However, we find that even with our adopted spatial selection, that there is a small population of what we believe are supergiants, lying along the blue end of this color band. To prevent the influence of these young stars on our measurements, we further restrict our color selection to within  F090W-F150W = 1.30$-$1.75 mag. That being said, we will later show that the precise color selection makes little difference to our measurements.

\subsection{Edge Detection Results}

The first group of measurement techniques involve edge-detection. To construct the underlying luminosity function, we first bin the photometry with a small bin-width ($\sigma_{bin}$= 0.01 mag), using only the stars in our limited color range. Before applying the edge-detector, we smooth the underlying luminosity function with a Gaussian locally-weighted regression smoothing (GLOESS) algorithm \citep{2004AJ....128.2239P, 2017ApJ...845..146H} to reduce the underlying noise present due to stochastic and/or star-formation history driven variations in the underlying stellar populations. Given that our color-magnitude diagram extends $\sim$3 magnitudes below the TRGB and that our CMD is relatively well populated even within our selected color band ($\sim$3200 stars in the first magnitude below the measured TRGB), we adopt a relatively small baseline smoothing scale of $\sigma_{S}$ = 0.05 mag. 

\begin{figure}
\epsscale{1.05}
\plotone{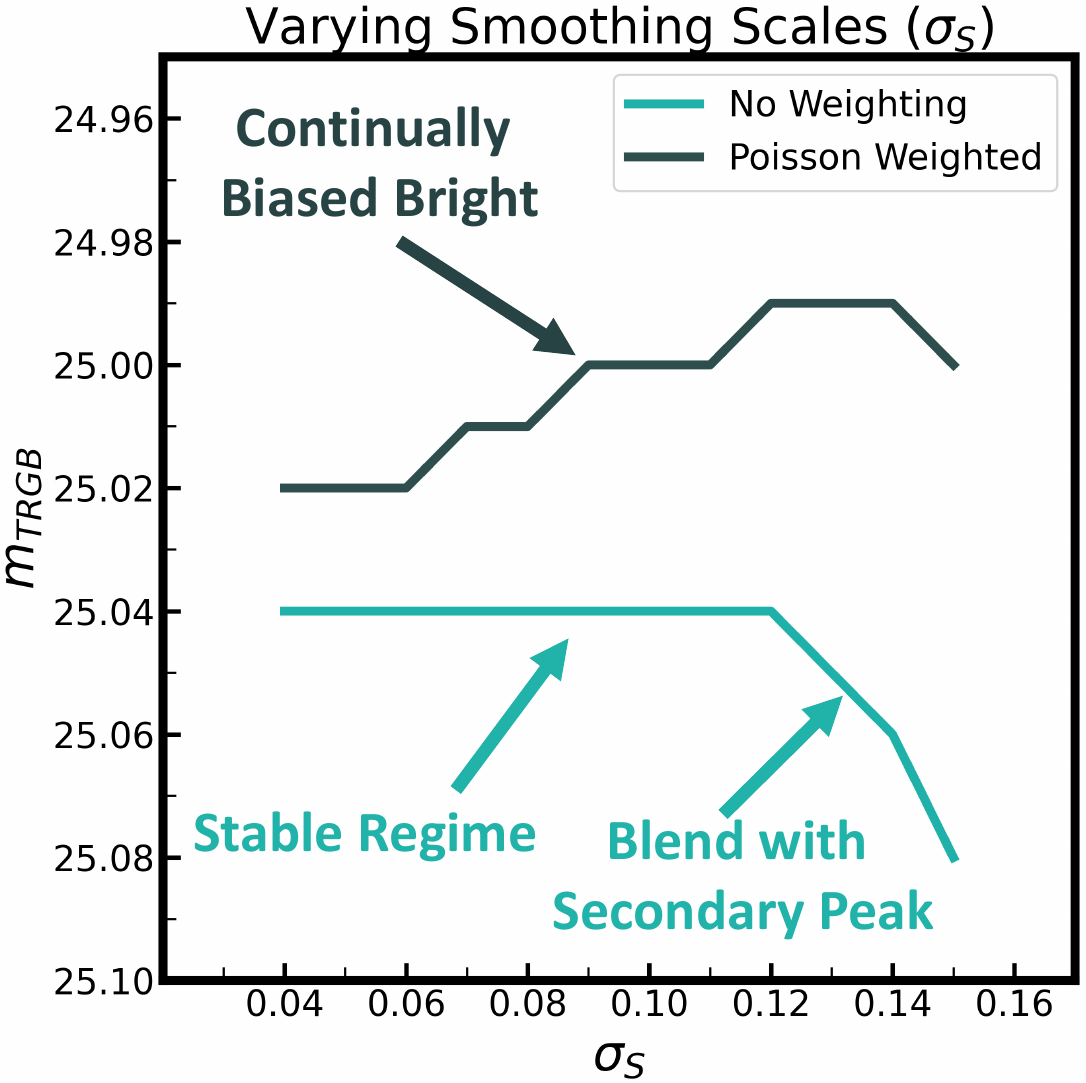}
\caption{The variation of $m_{TRGB}$ as a function of the underlying smoothing scale ($\sigma_{S}$), for both the unweighted and the Poisson-weighted Sobel edge detectors. The unweighted case shows a stable response until a relatively large smoothing scale, where the TRGB response begins to blend with a fainter edge response. The Poisson weighted response shows a general trend of becoming brighter with increase smoothing scales, with no readily apparent cause.}
\label{smoothing-scales}
\end{figure}

To perform the edge-detection, we employ the use of a Sobel filter with a kernel of [-1,0,1], which is effectively a discrete first derivative. Our baseline result is shown in Figure \ref{edge-detect}, from which we measure $m_{TRGB}$ = \baseEDR~$\pm$ \baseEDRerror~mag, where the uncertainty is determined via 1,000 bootstrap resampling with replacement trials, where each time the TRGB is remeasured (similar to \citealt{2018ApJ...868...96C, 2023NatAs...7..590H}). To test for any variations with the adopted value of $\sigma_{S}$, we run through a range of smoothing scales from 0.04$-$0.15 mag (see Figure \ref{smoothing-scales}), which bracket the values that are generally used in the literature. From $\sigma_{S}$~=~0.04$-$0.12 mag, we find no change in the measurement of $m_{TRGB}$. However, with smoothing scales higher than $\sigma_{S}$ = 0.12 mag, we find that the peak in the edge-detection response identified as the TRGB begins to blend with a second set of peaks near $m_{F090W}$$\sim$25.3 mag, causing the measurement to become skewed faint-ward due to a modest jump in the underlying luminosity function.

We note that our baseline result does not employ the use of further weighting to the edge-detection output, as done by some previous works \citep{2017ApJ...845..146H,2019ApJ...882...34F, 2023NatAs...7..590H}. We test the use of a Poisson-weighted weighting scheme over the same range of smoothing scales tested for the unweighted edge-detection. It can be seen in Figure \ref{smoothing-scales} that the weighted edge-detection results start becoming brighter with $\sigma_{S}$ $>$ 0.06 mag. This matches the behavior seen from simulated luminosity functions as well as the observed Large Magellanic Cloud red giant luminosity function \citep{Anderson2023}. Unlike the case with the unweighted edge-detection, there is no significant peak brightward of the one that is identified as the TRGB. Instead, we suspect this may be an inherent feature of weighted edge-detectors, and thus avoid utilizing them further. We recommend that future studies should closely examine the potential systematic effects of adopting different smoothing scales in their work, especially when employing a further weighting algorithm.

To test the effects of applying a cruder spatial selection, we also perform our measurement on the combined CMD generated from the outermost 2 of 8 SW NIRCam chips from each visit (the bold chips in Figure \ref{footprint}). We measure nearly the same value of $m_{TRGB}$ = \fourchipEDR $\pm$ \fourchipEDRerror~mag, indicating that our results are not highly sensitive to the precise spatial selection that is adopted. Indeed, \cite{2021ApJ...906..125J} found that their measured TRGB values in NGC 4258 did not significantly vary over a broad range of radial cuts (from 6$\arcsec>$ SMA $>$ 22$\arcsec$ in their mosaic). Subdividing the crude spatial cuts further into their individual visits results in $m_{TRGB}$ = 25.06 $\pm$ 0.05~mag for Visit 1 (East), and $m_{TRGB}$ = 25.04 $\pm$ 0.02~mag for Visit 2 (West). Again, our general point stands, in that our results are quite resilient to the precise spatial cut applied.

Additionally, we note that while we exercise additional care in our color selection (mainly to eliminate the contamination from a small sample of yellow supergiants), a much broader color range gives a very similar result$-$ that is, increasing the color range to 0.5$-$2.5 mag results in $m_{TRGB}$ = \broadcolorEDR~$\pm$ \broadcolorEDRerror~mag, which is only 0.01 mag offset from our baseline result. This difference is likely very small due to the fact that we do not have a large sample of higher metallicity TRGB stars in the outskirts of NGC~4258. These stars would appear faintward of the low-metallicity TRGB, and could otherwise skew our result towards substantially fainter magnitudes.

\subsection{Fitting a Model Luminosity Function}

\begin{figure}
\epsscale{1.05}
\plotone{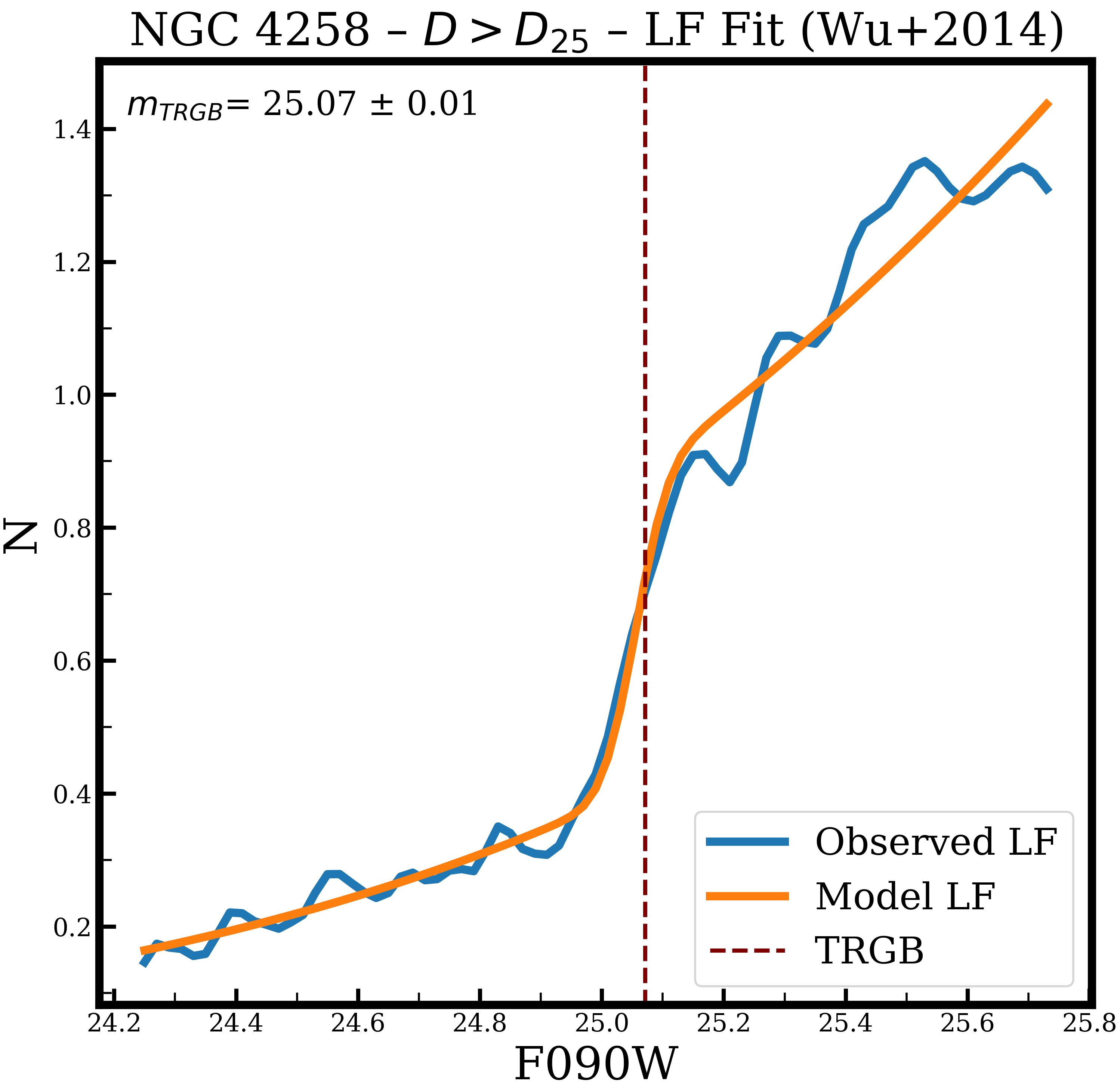}
\caption{The baseline model fit for the TRGB measurement via the method of \cite{2014AJ....148....7W}, from which we find $m_{TRGB}$ =~\baseTRGBTOOLWU~$\pm$~\baseTRGBTOOLWUerror~mag.}
\label{lf-fit}
\end{figure}

Next, we turn to the model-fitting solutions to determine the value of $m_{TRGB}$. We will explore the use of three different tools. First, using the original maximum likelihood \texttt{TRGBTOOL} described in \cite{2006AJ....132.2729M}, we find a value of $m_{TRGB}$ = \baseTRGBTOOL~$\pm$~\baseTRGBTOOLerror~mag. We also use the modified \texttt{TRGBTOOL} presented in \cite{2014AJ....148....7W}. As mentioned earlier, the underlying fitting algorithm in this work (a non-linear least squares Levenberg-Marquardt algorithm) differs from one utilized by \citet{2006AJ....132.2729M}, a BFGS maximum-likelihood algorithm. Notably, the tool described within \cite{2014AJ....148....7W} requires binning of the underlying luminosity function. The only change we make to the procedure as described within \cite{2014AJ....148....7W} is that we lower the bin-width from 0.05~mag to 0.02~mag, an option we are afforded due to the relatively well-populated CMDs. With this version of the tool, we find $m_{TRGB}$ = \baseTRGBTOOLWU~$\pm$ \baseTRGBTOOLWUerror~mag, where the error is determined from the square root of the variance of the TRGB magnitude. We also find that varying the bin-width between 0.01$-$0.05~mag in steps of 0.01~mag varies the measured TRGB magnitude by less than 0.005~mag. For illustration, we show the baseline model fit to the observed luminosity function using the \citep{2014AJ....148....7W} methodology in Figure \ref{lf-fit}. 

\begin{deluxetable*}{cccccc}[t]
\tabletypesize{\footnotesize}
\tablewidth{0pt}
\tablehead{
\colhead{Methodology} & \colhead{Color Range} & \colhead{Spatial Selection} & \colhead{$m_{TRGB}$ [mag]} & \colhead{$\pm$ [mag]}  & \colhead{Method Reference}
}
\startdata
Simple Sobel                & 1.30$-$1.75 & $D> D_{25}$ & \baseEDR               & \baseEDRerror                 & e.g. \cite{2019ApJ...882...34F} w/o weighting \\ 
LF Fitting                  & 1.30$-$1.75 & $D> D_{25}$ & \baseTRGBTOOL          & \baseTRGBTOOLerror            & e.g. \cite{2006AJ....132.2729M} \\ 
LF Fitting                  & 1.30$-$1.75 & $D> D_{25}$ & \baseTRGBTOOLWU        & \baseTRGBTOOLWUerror          & e.g. \cite{2014AJ....148....7W} \\
LF Fitting (w/o art. stars) & 1.30$-$1.75 & $D> D_{25}$ & \baseSeanLi            & \baseSeanLierror              & e.g. \cite{2022ApJ...939...96L} \\
\hline
\hline
Simple Sobel                & 1.30$-$1.75 & Outer Chips & \fourchipEDR           & \fourchipEDRerror             & e.g. \cite{2019ApJ...882...34F} w/o weighting \\
LF Fitting                  & 1.30$-$1.75 & Outer Chips & \fourchipTRGBTOOL      & \fourchipTRGBTOOLerror        & e.g. \cite{2006AJ....132.2729M} \\ 
LF Fitting                  & 1.30$-$1.75 & Outer Chips & \fourchipTRGBTOOLWU    & \fourchipTRGBTOOLWUerror      & e.g. \cite{2014AJ....148....7W} \\
LF Fitting \textbf{(w/o art. stars)}                 & 1.30$-$1.75 & Outer Chips & \fourchipSeanLi        & \fourchipSeanLierror          & e.g. \cite{2022ApJ...939...96L} \\
\hline
Simple Sobel                & 0.50$-$2.50 & $D> D_{25}$ & \broadcolorEDR         & \broadcolorEDRerror           & e.g. \cite{2019ApJ...882...34F} w/o weighting \\
LF Fitting                  & 0.50$-$2.50 & $D> D_{25}$ & \broadcolorTRGBTOOL    & \broadcolorTRGBTOOLerror      & e.g. \cite{2006AJ....132.2729M} \\ 
LF Fitting                  & 0.50$-$2.50 & $D> D_{25}$ & \broadcolorTRGBTOOLWU  & \broadcolorTRGBTOOLWUerror    & e.g. \cite{2014AJ....148....7W} \\
LF Fitting (w/o art. stars) & 0.50$-$2.50 & $D> D_{25}$ & \broadcolorSeanLi      & \broadcolorSeanLierror        & e.g. \cite{2022ApJ...939...96L} \\
\enddata
\caption{Table summarizing our four baseline TRGB measurements, followed by two variants for each (adjusting the spatial selection and color ranges). \label{tb:trgb}}
\end{deluxetable*}

Finally, we also use a maximum-likelihood algorithm based tool which does not employ the use of artificial stars. This method searches for the set of parameters in the broken power law luminosity function model that maximizes the likelihood of the sample, given the model and model parameters (the same underlying principle as \citealt{mendez2002}). A two-dimensional version of this tool is described in \cite{2022ApJ...939...96L}, where the extra dimension is required due to the need to jointly model individual red giants in the Milky Way that are not located at the same line-of-sight distance. We use a one-dimensional version here, and find $m_{TRGB}$ = \baseSeanLi~$\pm$ \baseSeanLierror~mag, where the error is determined from the inverse Hessian matrix. 

As with the edge-detection results, we also perform a series of the above three fits, but using a cruder spatial cut which involves just examining the outer two chips from each of the two visits. With the original~\texttt{TRGBTOOL} \citep{2006AJ....132.2729M}, we find $m_{TRGB}$ = \fourchipTRGBTOOL~$\pm$~\fourchipTRGBTOOLerror~mag. With the tool described within \cite{2014AJ....148....7W}, we find $m_{TRGB}$ = \fourchipTRGBTOOLWU~$\pm$~\fourchipTRGBTOOLWUerror~mag. Lastly, with a one-dimensional version of the maximum likelihood tool from \cite{2022ApJ...939...96L} (in which a two-dimensional version is required as Milky Way stars are not all at the same distance), we find $m_{TRGB}$ = \fourchipSeanLi~$\pm$ \fourchipSeanLierror~mag. The scattering of these results from their original values with the cruder spatial cut is cause for close examination of the adopted spatial selection criteria with future TRGB measurements. Although the differences are modest in our case (0.01$-$0.02 mag), these changes may be noticeably larger with data that is either lower S/N or taken in different regions of the host galaxy.

Lastly, we also test the effects of adopting a broader color baseline with our LF fitting techniques. We increase the color baseline from F090W$-$F150W = 1.30$-$1.75 mag to 0.5$-$2.5 mag. For the original \texttt{TRGBTOOL} \citep{2006AJ....132.2729M}, we now find $m_{TRGB}$ = \broadcolorTRGBTOOL~$\pm$ \broadcolorTRGBTOOLerror~mag, which is identical to the version with the stricter color selection (but with a larger uncertainty). With the modified \texttt{TRGBTOOL} \citep{2014AJ....148....7W}, we find $m_{TRGB}$ = \broadcolorTRGBTOOLWU~$\pm$ \broadcolorTRGBTOOLWUerror~mag, or 0.01 brighter than the baseline result with this methodology. Lastly, with the one-dimensional version of the \cite{2022ApJ...939...96L} tool, we find $m_{TRGB}$ = \broadcolorSeanLi~$\pm$ \broadcolorSeanLierror~mag, or 0.01 mag fainter than the baseline result. 

The modest (0.01$-$0.02~mag) differences we see with a much broader color selection for the LF fitting techniques are likely reflective of the fact that we simply do not have many high metallicity RGB stars in our spatially trimmed region. If working with a dataset with a substantial portion of high metallicity RGB stars near the tip, we anticipate our results would be skewed towards a fainter magnitude. This situation is similar to the same tests with a Sobel filter, where again we did not see a substantial difference when broadening the color range.

\subsection{Differences between Measurement Routines}

We provide a summary of all of our measurement techniques and their resultant TRGB magnitudes in Table \ref{tb:trgb}. The table includes the four baseline results for each distinct measurement methodology, as well as two variants for each (one which varies the color selection, and a second which varies the spatial selection).

\begin{deluxetable*}{lccc}[t]
\tabletypesize{\small}
\tablewidth{0pt}
\tablehead{
\colhead{Uncertainty} & \colhead{Result} [mag] & \colhead{$\sigma_{stat}$ [mag]} & \colhead{$\sigma_{sys}$ [mag]} 
}
\startdata
TRGB Measurement & 25.055 & 0.02 & 0.02 \\
Intrinsic TRGB Variation & -- & -- & 0.02 \\
Color Selection & -- & 0.01 & -- \\
Spatial Selection & -- & 0.01 & -- \\
WebbPSF Model & -- & 0.02 & -- \\
NIRCam PSF Stability & -- & 0.01 & -- \\
NIRCam Zeropoints & -- & -- & 0.02* \\
DOLPHOT Photometry & -- & -- & 0.03* \\
\hline
\hline
$m_{TRGB}^{F090W}$ & 25.055 & 0.033 & 0.028 \\
$A_{F090W}$ & 0.0198 & -- & 0.014 \\
$m_{TRGB}^{F090W,0}$ & 25.035 & -- & -- \\
\hline
Maser Distance Modulus & 29.397 & -- & 0.032 \\
\hline
\hline
$M_{TRGB}^{F090W}$ & $-$4.362 & 0.033 & 0.045 \\
\hline
\hline
\enddata
\caption{Uncertainty budget for our absolute calibration of $M_{TRGB}^{F090W}$. For the foreground extinction, we assign a conservative systematic uncertainty of 0.014~mag, which is the quadrature sum of one-half of the measured extinction and an additional 0.01~mag of uncertainty due to potential extinction within the outer regions of the host galaxy itself. \textbf{*Note:} The uncertainty terms assigned to the NIRCam Zeropoints and DOLPHOT photometry will cancel when comparing photometry which uses the same set of NIRCam zeropoints and reduction setup, and so we do not include these uncertainties in our totals.\label{tb:calib}}
\end{deluxetable*}

It is interesting to note that the baseline results could be grouped into two categories, with one being brighter (25.04 mag) than the other (25.07 mag). However, the split is not upon the edge-detection versus LF-fitting methodologies, but instead on whether or not artificial stars are included in the analysis. We speculate that this difference of 0.03 mag could arise from the fact that the photometric error of stars near the tip are 0.04 mag in F090W (see Figure \ref{curves}), similar to the level offset seen between the results (whereas the photometric bias is $<$0.01 mag). It is possible that the techniques that do not rely on artificial stars are triggering off a set of up-scattered TRGB stars (whereas those that do use artificial stars take this effect into account in their modelling of the luminosity functions). This effect could explain offsets between TRGB measurements by different groups of the same galaxies in the literature \citep{2021ApJ...906..125J, 2022ApJ...932...15A}. On the other hand, this could be reading too much into our results, and the differences may just arise from underlying uncertainties in the various fitting procedures, which may under-report the true uncertainties. Yet another alternative is that the artificial star procedures need further refinement, with the minor NIRCam exposure time issues (see $\S$3 for a brief discussion) coming into play. Future, more detailed investigations are warranted. 

Broadly speaking, methodological differences in TRGB measurements for the same underlying data at the level of $\sim$0.04 mag are common in the literature on a case-by-case basis \citep{2022ApJ...932...15A}. Avoiding a systematic uncertainty between measurements at this level is difficult without substantial effort to match photometric reduction techniques, noise properties, populations via selection, TRGB measurements techniques, etc. While an error at this level is less relevant for a common galaxy, it is most significant in the case of NGC~4258, which serves as a calibrator of many.

\section{Absolute Calibration}\label{sec:results}

To provide our absolute calibration of the TRGB in F090W, we begin with our value of $m_{TRGB}$ = 25.055 $\pm$ 0.02~mag, where the measurement is the simple average of our four baseline techniques and the uncertainty is the larger of the reported values (two methods provide 0.01~mag, and two provide 0.02~mag). To proceed further, we must first determine the amount of foreground extinction caused by the Milky Way. We used the \citet{1999PASP..111...63F} reddening law with $R_V$ = 3.1, convolved with the transmission functions of the Johnson $B$, Johnson $V$, F090W, and F150W filters, to determine the effective reddening relationships between these filters. This analysis yielded reddening coefficients of $A_{F090W}/E(B-V)$ = 1.4156 and $A_{F150W}/E(B-V)$ = 0.6021. Thus, from E(B-V) = 0.014 \citep{2011ApJ...737..103S}, we determine $A_{F090W}$ = 0.0198~mag and $A_{F150W}$ = 0.0084~mag. Correcting our measured value for foreground extinction, we find $m_{TRGB}^{F090W,0}$ = 25.035~mag. The value for the geometrical distance modulus to NGC~4258 provided by \cite{2019ApJ...886L..27R} is $\mu$= 29.397 $\pm$ 0.032 mag, or D = 7.576 $\pm$ 0.082 (stat.) $\pm$ 0.076 (sys.) Mpc. Subtracting the two distance moduli, we find $M_{TRGB}^{F090W}$ = $-$4.362~mag. 

\begin{figure*}
\epsscale{1.125}
\plotone{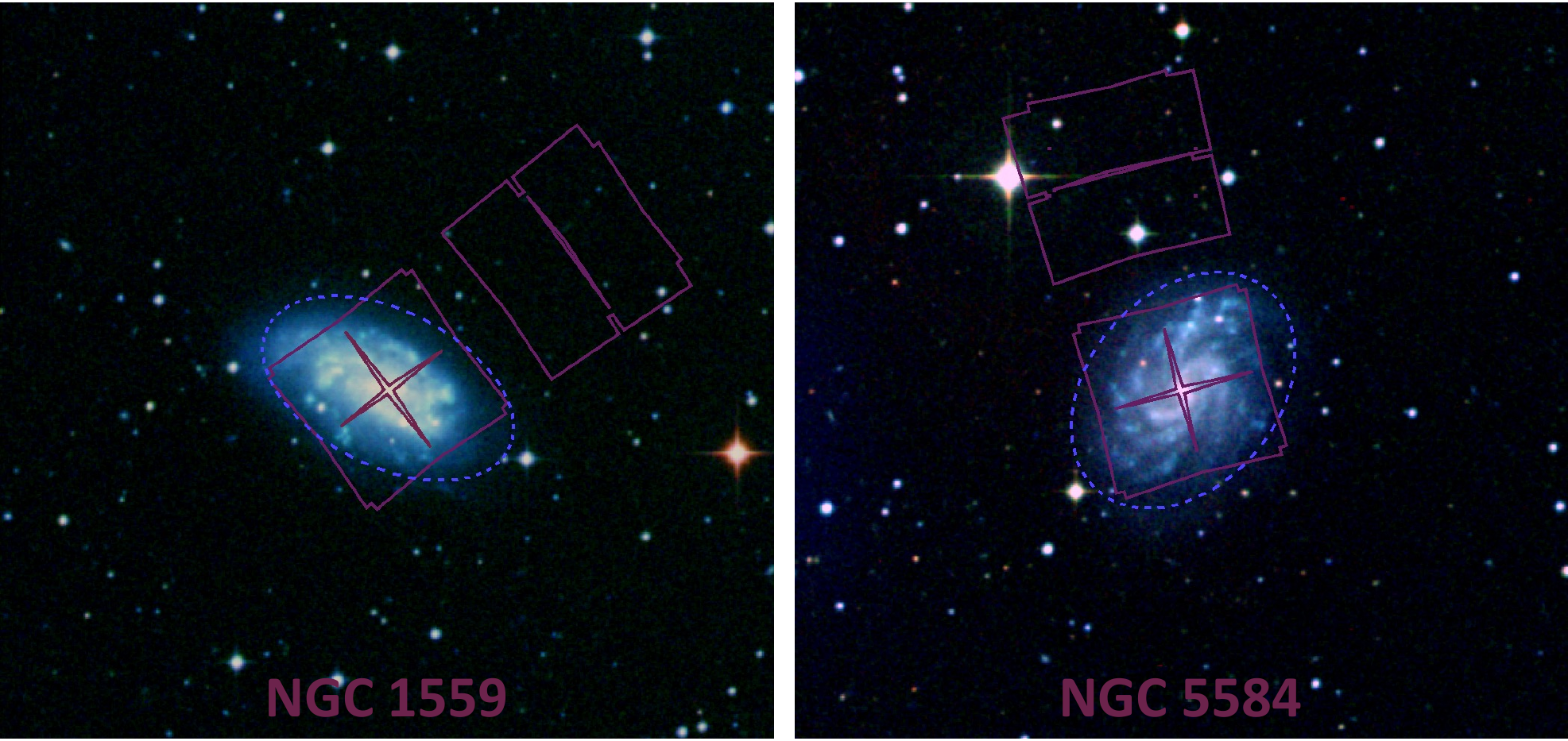}
\caption{Footprints for our NGC~1559 and NGC~5584 visits, overlaid on Digitized Sky Survey color images. We use only the outer ``halo" modules for our TRGB analysis, as they are well removed from the spiral disks and lie outside $D_{25}$ (shown as the dashed blue lines).}
\label{sn-footprints}
\end{figure*}

Now, onto the important matter of the uncertainty associated with our calibration. We lay out the separate sources of uncertainty in Table \ref{tb:calib}, as done similarly in Table 5 of \cite{2021ApJ...906..125J}. The individual uncertainties are separated by whether or not they could be reduced with additional observations around NGC~4258 (i.e. statistical vs. systematic). For instance, the initial TRGB measurement is assigned a statistical uncertainty of 0.02~mag (as mentioned at the beginning of this subsection), and an additional systematic uncertainty of 0.02~mag (derived from the spread in our four baseline measurements). A further 0.02~mag systematic contribution is assigned due to the potential variations of $M_{TRGB}^{F090W}$ with age and metallicity, which are small but non-zero (see Figure \ref{parsec}). This uncertainty term may also encompass some of the differences seen in recent field-to-field measurements done by the CATs team \citep{2023ApJ...954...87W}, though our TRGB measurement techniques differ from theirs, complicating a more direct comparison. Our color and spatial selections are assigned uncertainties of 0.01~mag each, based on the impact of varying these quantities (shown in Table \ref{tb:trgb}). 

Moving to uncertainties resulting from our photometry, we note the presence of small mismatches between the WebbPSF models \citep{2014SPIE.9143E..3XP} currently used within DOLPHOT and the observed PSFs, which range between 0 and 2$\%$ in the central pixel of the PSF (in either positive and negative directions, depending upon the particular frame in question). While the PSF models are empirically varied by DOLPHOT on a frame-by-frame basis, we still choose to include a 0.02 mag uncertainty term here. Another small contribution is derived from the stability of the NIRCam PSF, which \cite{2023ApJ...956L..18R} show to be very consistent for the observations in question. One sometimes overlooked uncertainty contribution is the difference between a given photometry setup (e.g. DOLPHOT's NIRCam module with recommended parameters), versus other reductions for which our absolute calibration may be applied to (e.g. with different DOLPHOT setup parameters, or using other photometry packages). We adopt a 0.03 mag uncertainty term here, based on the extensive tests carried out in \cite{2021ApJ...906..125J} and \cite{2023MNRAS.521.1532J}.

Lastly, for the foreground extinction, we assign a conservative systematic uncertainty of 0.014~mag, which is the quadrature sum of one-half of the measured extinction and an additional 0.01~mag of uncertainty due to potential extinction within the outer regions of the host galaxy itself (as also done by \citealt{2021ApJ...906..125J}). In the end, we find 
\begin{equation*}
\bm{M_{TRGB}^{F090W} = -4.362~\pm~0.033~\mathrm{(stat)}~\pm~0.045~\mathrm{(sys)~mag}}
\end{equation*}
when excluding the error contributions which cancel with comparisons to photometry taken with the same set of NIRCam zeropoints, and reduced in an identical manner with DOLPHOT. If including these terms, the systematic uncertainty increases from \calibsys~mag to 0.058~mag.

We note that while we provide one averaged calibration for our four underlying methods, this can be further tailored to match a particular TRGB measurement methodology. For instance, if one were using the precise methodology outlined in \cite{2006AJ....132.2729M}, a value of $M_{TRGB}^{F090W}$ = $-$4.347~mag may be more appropriate. Similarly, $M_{TRGB}^{F090W}$ = $-$4.377~mag may be appropriate for providing the absolute calibration to an unweighted Sobel measurement (as we use for NGC~1559 and NGC~5584 in the next section).

\begin{figure*}[t!]
\epsscale{1.1}
\plotone{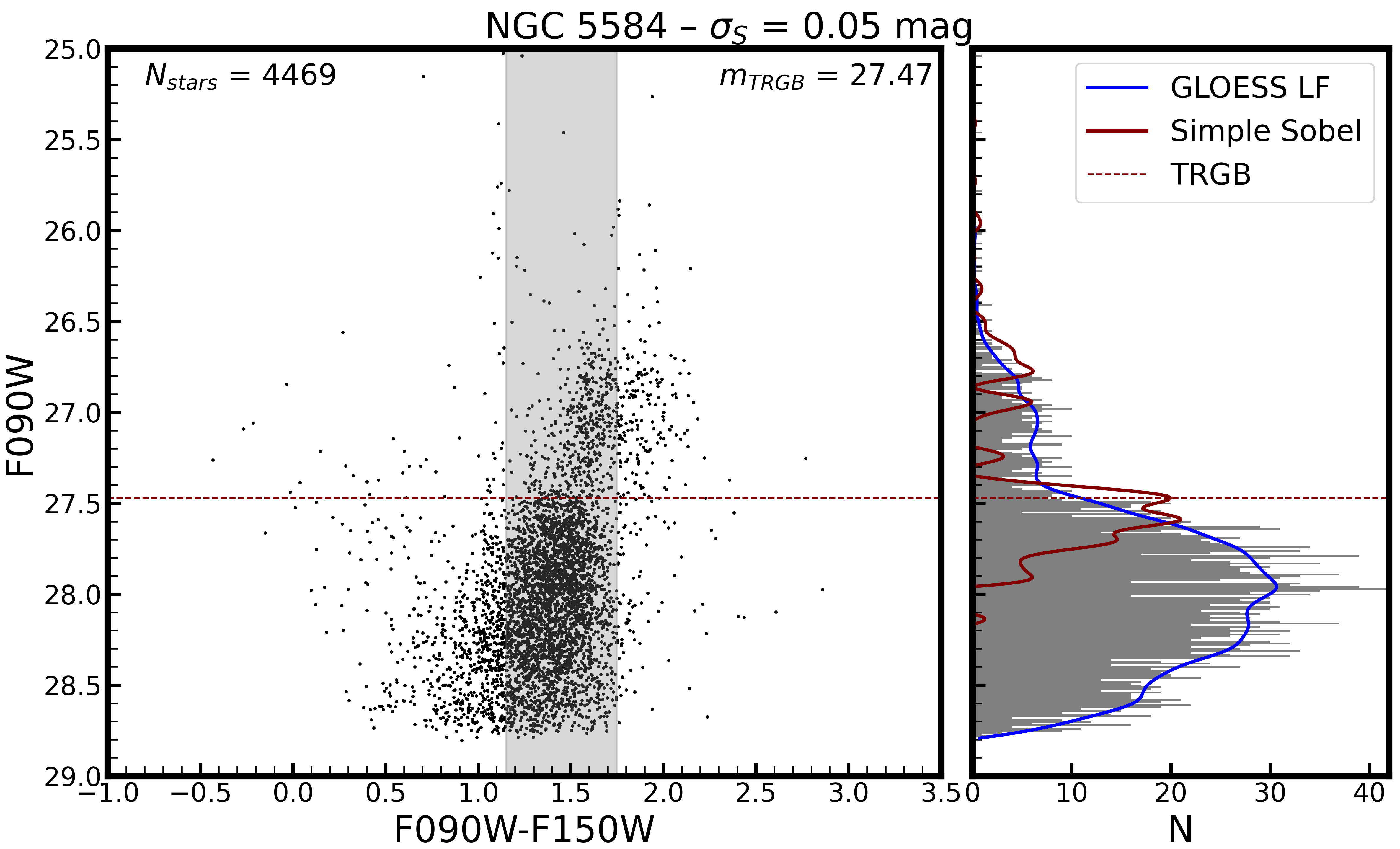}
\caption{Our simple Sobel TRGB measurement for NGC~5584, performed in the same manner as was done for our baseline edge detection of NGC~4258 (shown in Figure \ref{edge-detect}). We use only the outer ``halo" module for our TRGB analysis (see Figure 3 in \citealt{2023ApJ...956L..18R}), as it is well removed from the spiral disk. The TRGB is clearly visible, even to the eye, with a sharp contrast between the AGB above (which ``fans" off to the right), and the RGB below.}
\label{n5584}
\end{figure*}

\section{Application to Type Ia Supernova Hosts}

With an initial calibration in place, we are able to determine absolute distances to other galaxies observed with JWST in the F090W+F150W filter set. Here, we will highlight the cases of NGC~1559 and NGC~5584, which are host to the type Ia supernovae SN2005df \citep{2005CBET..192....1E} and SN2007af \citep{2007IAUC.8817....3N}, respectively.

\subsection{NGC~5584}

NGC~5584 was observed as part of the same JWST program as NGC~4258, with the main difference being greater exposure times, and the field placement. In particular, the outer NIRCam module placements from each visit are mostly overlapping, with a minor orientation difference between the two (see the right-hand side of Figure \ref{sn-footprints}). This decision was made to provide greater depth for TRGB measurements, given that the Cepheid distance to this galaxy is $\sim$23 Mpc \citep{2022ApJ...934L...7R}. As mentioned in \cite{2023ApJ...956L..18R}, NGC~5584 is an important galaxy for examining the Hubble tension, as it (1) is near the mean distance of the SH0ES galaxies, (2) contains a plethora of known Cepheids, and (3) on its own, as a singular target of comparison, is nearly 3$\sigma$ inconsistent with $H_0 \sim 67.5$ km/s/Mpc based on results from the {\it Planck} satellite \citep{2020A&A...641A...6P}. 

A TRGB distance to NGC~5584 was first presented by \cite{2015ApJ...807..133J}, who found $\mu$ = 31.76 $\pm$ 0.04 (random) $\pm$ 0.12 (systematic) mag, which was measured in the outskirts of the spiral disk using HST data acquired to measure Cepheids. A re-scaled version of this same TRGB measurement was utilized by \cite{2019ApJ...882...34F} in determining a value of the Hubble constant from the TRGB (instead of Cepheids). However, a re-analysis of the same underlying images by \cite{2022ApJ...932...15A} was unable to measure the TRGB, due to the shallowness of the resulting photometric catalog. \cite{2023ApJ...954L..31S} came to the same conclusion, using the same photometric catalog produced by \cite{2022ApJ...932...15A}. Now, we perform an analysis of the TRGB from this Cycle 1 JWST dataset, which shows a clearly visible presentation of the TRGB. Recently, \cite{2023ApJ...956L..18R} determined a revised Cepheid distance with the B modules (placed on the disk) from this same JWST dataset, where they measure a baseline distance modulus of $\mu$ = 31.813 $\pm$ 0.020 mag. 

\begin{figure*}
\epsscale{1.15}
\plottwo{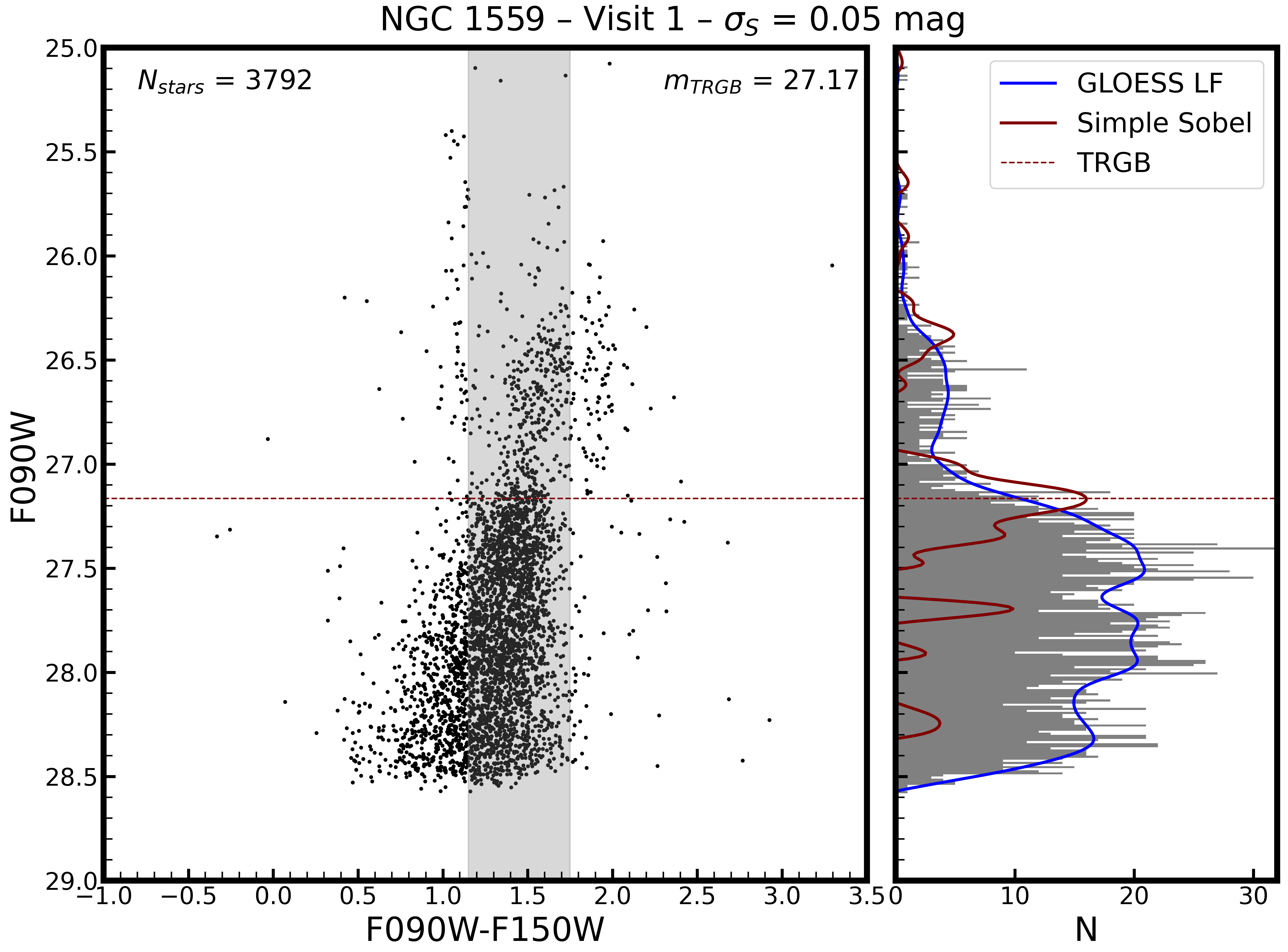}{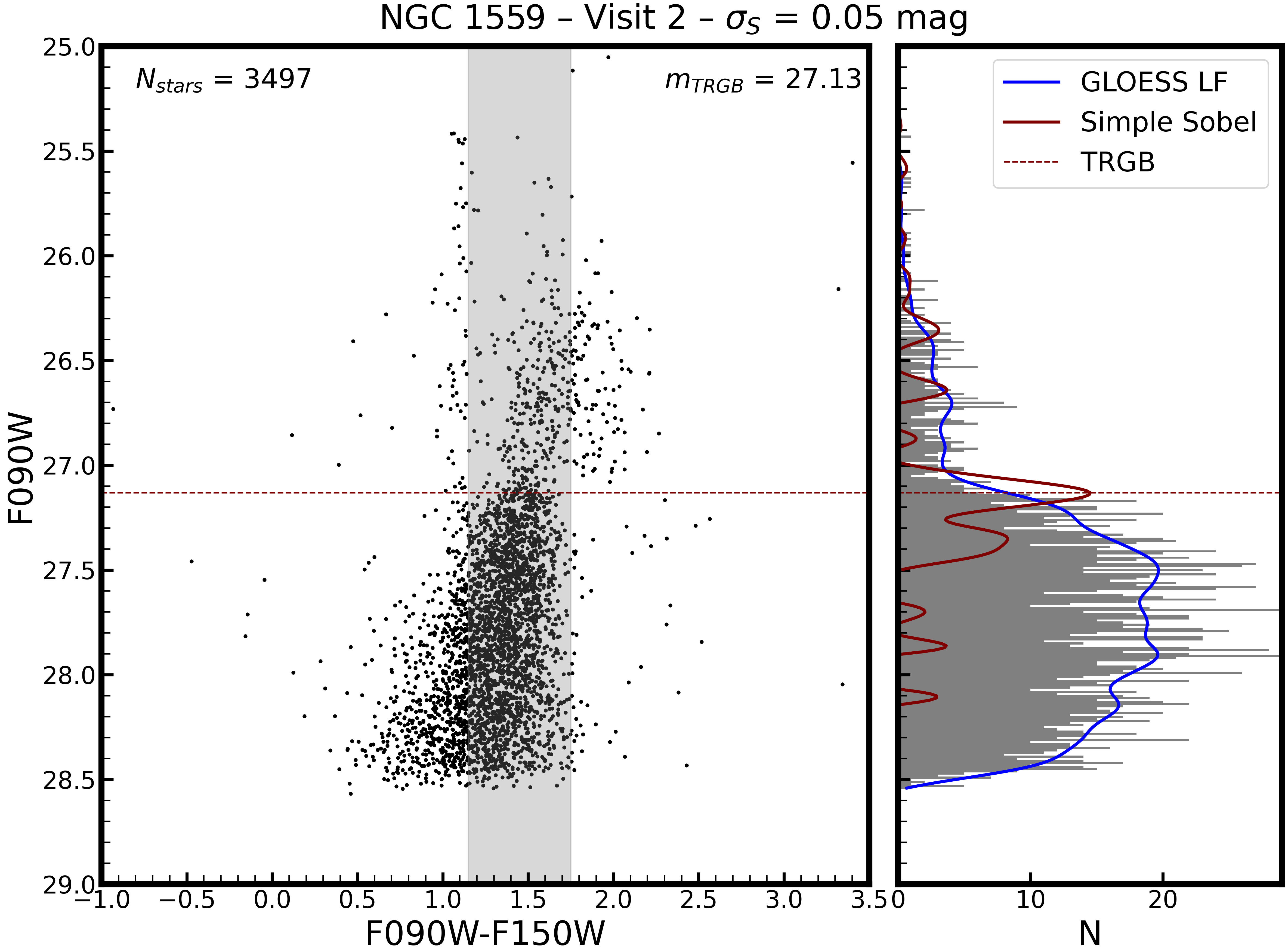}
\caption{The two panels show our simple Sobel TRGB measurements for each visit of NGC~1559, performed in the same manner as was done for our baseline edge detection of NGC~4258 (shown in Figure \ref{edge-detect}). Again, we use only the outer ``halo" module for our TRGB analysis (see Figure \ref{sn-footprints}), as it is well removed from the spiral disk.}
\label{n1559-trgb}
\end{figure*}

We perform the reduction and analysis for NGC~5584 in the same manner as was done with NGC~4258, except that we only use the inner two chips of the A module, which fall within the target's halo and outside $D_{25}$ (the outer two chips of the A module are too sparsely populated). We also re-extend the left-edge of the color baseline to F090W$-$F150W = 1.15 due to the lack of younger supergiant stars (this alteration impacts the final result by only 0.01 mag). The resulting CMD is shown in Figure \ref{n5584}. We assign the first significant edge detection response to the TRGB, the most common choice in the literature and consistent with the analysis of NGC 4258, or $m_{TRGB}$ = \snhostEDR~$\pm$~\snhostEDRerror~mag. We note there is a substantial secondary response $\sim$0.1 mag fainter. This is not entirely surprising, as the data is of much lower S/N than it is for NGC~4258, and that the underlying CMD is of a sparser nature. Very similar secondary responses have been seen in the literature (see the case of NGC 1316 presented in \citealt{2018ApJ...866..145H}), and may also be in part due to variations in the underlying star formation history, assemblies of the stellar halos in question, or other factors. 

Accounting for $A_{F090W}$ = 0.05 mag \citep{2011ApJ...737..103S}, and using the TRGB magnitude from the analogous measurement in NGC~4258 (a simple Sobel filter), we determine $\mu$ = 31.80 $\pm$ 0.08~mag, in very good agreement with the recently determined JWST Cepheid distance of $\mu$ = 31.813 $\pm$ 0.020 mag \citep{2023ApJ...956L..18R} and the HST Cepheid distance of $\mu$ = 31.810 $\pm$ 0.047 mag \citep{Javanmardi2021}.

While the current situation with the JWST exposure times in image headers prevent the determination of highly precise estimates of S/N for individual stars (in the form of photometric errors), we can still use artificial stars to measure the impact of photometric \textit{bias} on our measurements. The increased exposure times and even lower stellar density levels (when compared to NGC~4258) result in a negligible photometric bias ($<$0.01~mag) at the magnitude of the TRGB, despite it being notably fainter than in NGC~4258. We anticipate a more rigorous analysis of the data from this JWST program (including results from other measurement routines), along with the other supernova hosts from this program (NGC~1448, NGC~1559, and NGC~5643) in a planned future work (Li et al., in prep).

As an entirely distinct point, we note that there is perhaps a visible gap between the brightest RGB stars, and the TP$-$AGB sequence above it (shown as an underdensity of stars in the first couple tenths of a magnitude brighter than the TRGB). This gap is predicted by some stellar models, whereas it is not in others. Here we simply note the appearance of this potential gap in our data, and highlight that it may be used to calibrate stellar models, as previously suggested by \cite{2012ApJS..198....6D}.

\subsection{NGC 1559}

Along with NGC~5584, we also present a TRGB measurement for NGC~1559. We select NGC~1559 as a second highlight because it is the only other galaxy from our program which does not have a TRGB measurement in the literature\footnote{Neither does NGC~5468, but at a distance of $\sim$40~Mpc, our dataset is too shallow to allow for such a measurement.}. Additionally, NGC~1559 has an independently determined distance modulus from its Mira variable stars of $\mu$ = 31.41~$\pm$~0.05 (stat.)~$\pm$~0.06 (sys.) mag \citep{2020ApJ...889....5H}, which allows for another avenue of comparison.

While we have the same two visit structure for NGC~1559, this galaxy is $\sim$0.4 mag closer than NGC~5584, allowing us to securely measure the TRGB from each visit separately. While there is a modest orientation difference between the two visits (see the left-hand side of Figure \ref{sn-footprints}), much of the underlying stellar population is the same, and a second measurement allows for an additional check of internal stability. 

\begin{figure*}
\epsscale{1.15}
\plotone{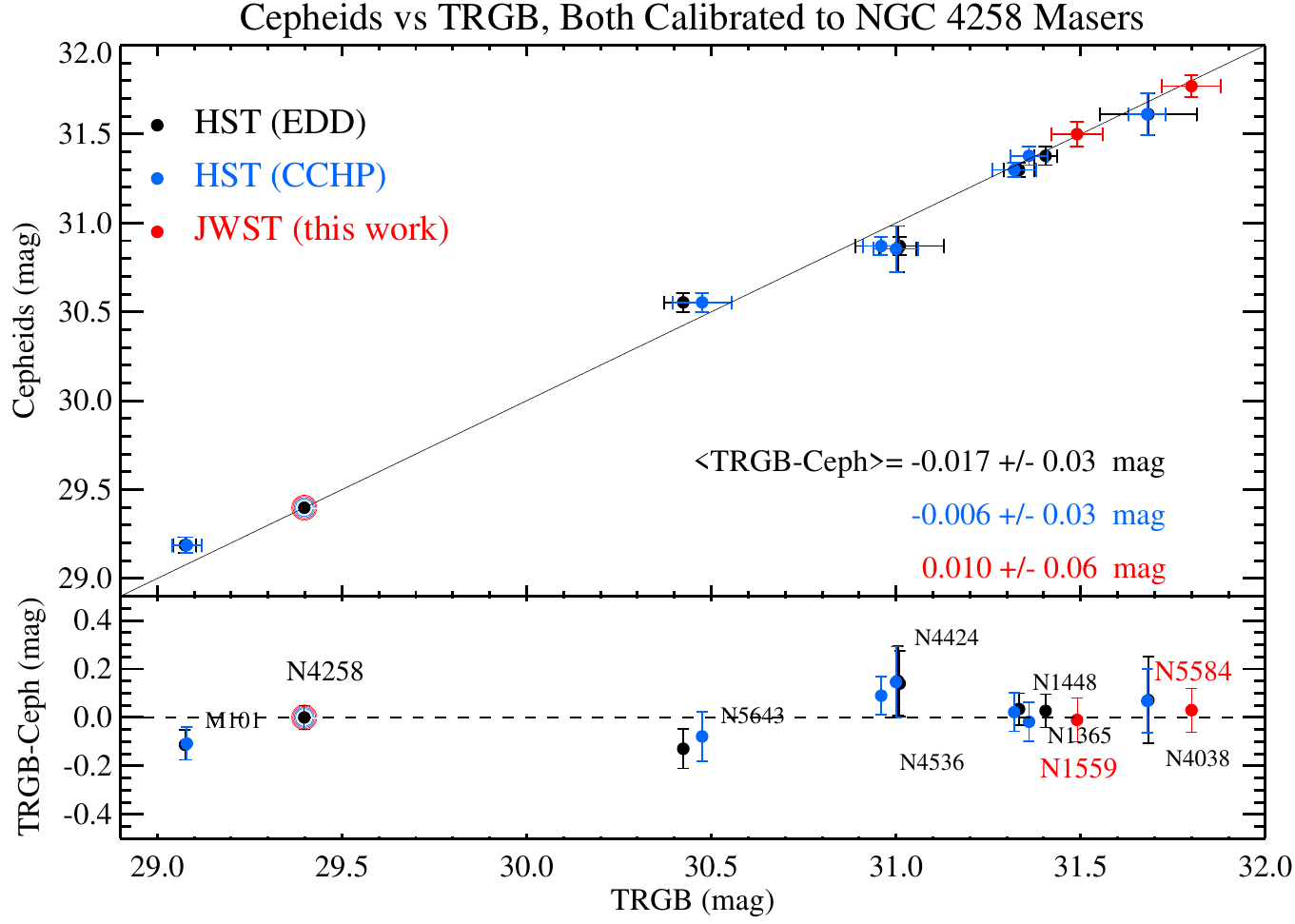}
\caption{An updated version of Figure 23 from \cite{2022ApJ...934L...7R}, which compares Cepheid and TRGB distances to Type Ia supernova host galaxies. We find excellent agreement between Cepheid and TRGB distances when each set is measured consistently relative to the same anchor (NGC~4258). }
\label{trgb-cepheid}
\end{figure*}

We follow the same procedure as described for NGC~5584, including the color baseline of F090W$-$F150W = 1.15$-$1.75. For our first visit, we find $m_{TRGB}$ = \mirahostvisitoneEDR~$\pm$~\mirahostvisitoneEDRerror~mag, and for the second visit we find $m_{TRGB}$ = \mirahostvisittwoEDR~$\pm$~\mirahostvisittwoEDRerror~mag. As in NGC 5584 and NGC 4258 we note the presence of a fainter feature, while in all three cases accepting the brighter.  Taking the simple average of this two measurements (and conservatively adopting the larger error), we find $m_{TRGB}$ = 27.15 $\pm$ 0.04~mag. From \cite{2011ApJ...737..103S}, we determine $A_{F090W}$ = 0.04~mag, and thus we find $\mu$ = 31.49 $\pm$ 0.07~mag. This is consistent with both the Cepheid determination of 31.49 $\pm$ 0.06~mag \citep{2022ApJ...934L...7R}, and the Mira determination of 31.41 $\pm$ 0.08~mag \citep{2020ApJ...889....5H}. As with NGC~5584, the photometric bias is negligible at the magnitude of the TRGB, and we anticipate a more multi-faceted and thorough analysis of this dataset in a future paper.

\subsection{Comparing Cepheid and TRGB Distances}

An important issue under study in the context of the current Hubble tension is the level of agreement between values of the Hubble constant derived from SN Ia calibrated through either Cepheids, TRGB and Mira distances as well as from Masers, Surface Brightness Fluctuations, and the Tully-Fisher relation, independent of SNe Ia. The Cepheid-SN Ia route gives a result of $H_{0} =$ 73.04 $\pm$ 1.04~km/s/Mpc with the most recent major SH0ES release \citep{2022ApJ...934L...7R}, with more recent updates \citep{2022ApJ...938...36R, 2023JCAP...11..046M} providing a value of 73.29 $\pm$ 0.90~km/s/Mpc.  

A prominent measurement of $H_{0}$ was derived via a TRGB calibration of Type Ia supernovae luminosities by the CCHP team \citep{2019ApJ...882...34F}, who found $H_{0} =$ 69.8 $\pm$ 0.8 (stat) $\pm$ 1.7 (sys)~km/s/Mpc. \cite{2021ApJ...919...16F} then provide an updated value of $H_{0} =$ 69.8 $\pm$ 0.6 (stat) $\pm$ 1.6 (sys)~km/s/Mpc. An alternate reduction from the same underlying HST data but with different photometric and TRGB measurement methodologies by \cite{2022ApJ...932...15A} resulted in $H_{0} =$ 71.5 $\pm$ 1.5~km/s/Mpc. Another alternate measurement by \cite{2023ApJ...954L..31S}, which uses the same photometry as \cite{2022ApJ...932...15A} from the Extragalactic Distance Database (EDD; \citealt{2009AJ....138..323T,2021AJ....162...80A}) but a different approach to standardizing TRGB measurements, finds a baseline result of $H_{0} =$ 73.22 $\pm$ 2.06~km/s/Mpc.

\begin{figure*}
\epsscale{1.15}
\plotone{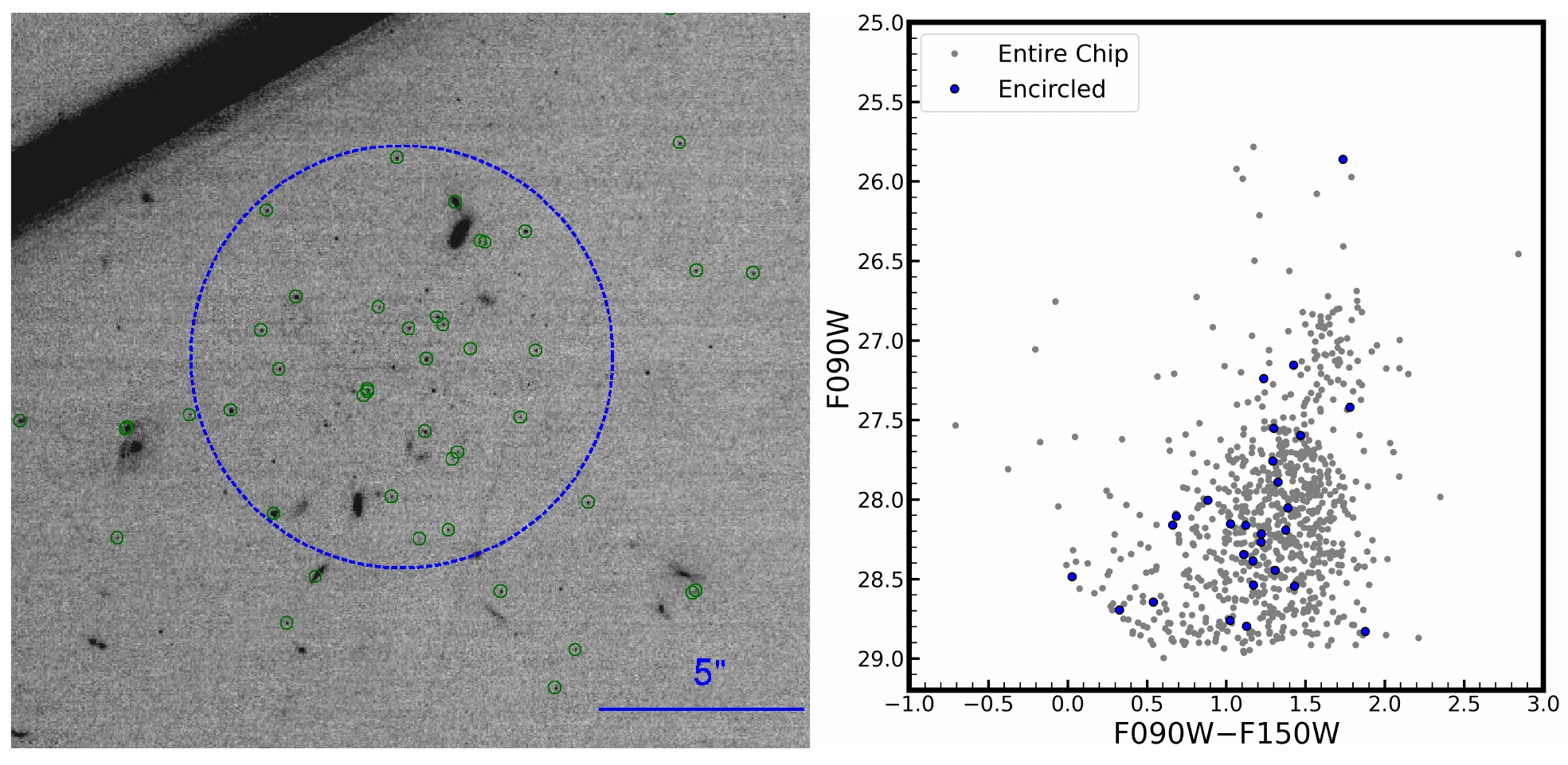}
\caption{The left-hand panel shows a cutout from the A3 chip of the first visit in NGC~5584. Encircled in blue is a candidate dwarf galaxy (NGC~5584$-$dw1) we identified upon first inspection of the images, with sources that pass our quality cuts encircled in green. The color-magnitude diagram within the circle is shown on the right in blue points, with the underlying CMD of the rest of the A3 chip shown in grey.}
\label{dwarf}
\end{figure*}

As discussed by \cite{2023ApJ...954L..31S}, the largest source of differences between the TRGB and Cepheid results are not due to differences between the use of Cepheids and TRGB to measure the distances to SN Ia host galaxies. Rather, they find in their comparisons of distance ladders that the majority of the disagreement between the $H_{0}$ values between these studies comes from differences in their treatment of the Type Ia supernova data in the rung following either Cepheids or TRGB. Specifically, the application of peculiar velocity corrections and photometric standardization of different SN surveys at optical wavelengths contributes to a $\sim$2.0 km/s/Mpc difference between the CCHP and studies which utilize the Pantheon+ SN survey \citep{2022ApJ...938..111B}. A recent study from \cite{2023arXiv230801875U} which combines Cepheids, TRGB, SN Ia, and SBF and includes application of SN Ia peculiar velocity corrections finds $H_{0} =$ 72.5 $\pm$ 1.5 ~km/s/Mpc.

For a direct comparison of TRGB and Cepheid measures as they contribute to the Hubble Tension, we can circumvent SNe Ia and geometric distance estimates and directly compare distances to individual SN Ia host galaxies determined with each method calibrated with the same geometric source, as is done in Figures 5 and 6 in \cite{2019ApJ...882...34F} and Figure 23 in \cite{2022ApJ...934L...7R}. Here, we provide an updated version of Figure 23 from \cite{2022ApJ...934L...7R} in our Figure \ref{trgb-cepheid}, which incorporates our two initial distance measurements to NGC~1559 and NGC~5584. The difference between the HST Cepheid and JWST TRGB distances when calibrated to the same anchor (NGC~4258) is negligible at 0.01 $\pm$ 0.06~mag, similar to the minor differences found between comparisons of HST Cepheids and HST TRGB results from the CCHP and EDD teams. With newly obtained and further upcoming JWST data that is appropriate for both Cepheid and TRGB measurements from GO$-$1685 \citep{1685prop}, GO$-$1995 \citep{2021jwst.prop.1995F}, and GO$-$2875 \citep{2023jwst.prop.2875R}, we anticipate even more robust comparisons between the Cepheids and TRGB in the near future. Comparisons of distances by different methods provide the ability to disentangle components of the distance ladder without reference to $H_0$ whose calculation is a composite of many steps.  For reference, the approximate size of the Hubble Tension is $5\log(73/67.5)=0.17$~mag, a difference which is inconsistent with the comparisons presented here, demonstrating that the middle rung of the distance ladder cannot be solely responsible for the Tension.

\subsection{Discovery of a New Dwarf Galaxy near NGC~5584}

While inspecting the images from the first visit of NGC~5584, we noticed what appeared to be a stellar overdensity on the A3 chip that was well off of the main spiral disk (which was focused on with the B module). We consider this stellar overdensity (see the left-hand side of Figure \ref{dwarf}) to be a candidate dwarf galaxy in the vicinity of NGC~5584.

While there are two visits on this galaxy, the orientation of the second visit misses half of the body of the dwarf candidate. Additionally, a significant portion of the remaining stars that are covered in the second visit lie in the glow of a diffraction spike of a very bright foreground star, which would likely lead to degraded photometry. For this reason, we only present data from the first visit for this newly discovered candidate dwarf galaxy, NGC~5584$-$dw1. The color-magnitude diagram of this dwarf galaxy is shown on the right-hand side of Figure \ref{dwarf}, limited to the region within the blue circle. The CMD for the entire chip from that same visit is shown in the background in grey points. We reduce the S/N cut to 3 instead of our baseline value of 5 for both filters, in order to probe further down the red giant branch. It can be seen that the candidate dwarf's CMD is broadly consistent with the underlying CMD from the halo of NGC~5584, with a somewhat lower mean metallicity. We reserve a more detailed analysis of the properties of this newly discovered dwarf galaxy, along with one more found in the outskirts of NGC~5468, to a future paper.


\section{Summary and Future Outlook} \label{sec:summary}

We present an initial, absolute calibration of the magnitude of the tip of the red giant branch in the JWST F090W passband based on data taken in the outer regions of the megamaser host NGC~4258. We find $M_{TRGB}^{F090W}$ = \calib~$\pm$ \calibstat~(stat) $\pm$ \calibsys~(sys)~mag, when considering only the relatively metal-poor stars before the high-metallicity ``turnover". The relative constancy of the TRGB in F090W at lower metallicities (traced by the color of stars) allows it to be used as a standard candle without the need for color ``rectification", as would be required with observations in bluer or redder filters. While we do not present this as a definitive calibration, we believe this work represents a first major step, and that any remaining systematic errors due to field placement ($<$ 0.02 mag) and uncertainties in the underlying photometric calibrations ($<$ 0.02 mag) are negligible for most applications (especially as the later would cancel out, when comparing data which uses the same version of NIRCam zeropoints).

Our NGC~4258 data, which consists of just 1030 seconds of exposure time in F090W at each field placement, would allow us to measure distances out to $\sim$19 Mpc (while keeping the TRGB one full magnitude above the S/N = 5 floor). A similar amount of time in F150W would suffice for the TRGB (the F150W dataset presented here is somewhat deeper, as it was required for the Cepheid observations). Including overheads, such a visit typically requires less than two hours of JWST charged time. This situation represents nearly a factor of 8 increase in volume over a similar setup with HST/ACS (a reach of 10~Mpc with 1 orbit split between F606W+F814W), highlighting the gains due to JWST's increased sensitivity. While largely not relevant to the measurement of the TRGB within NGC~4258, the increased resolution of NIRCam (compared to ACS/WFC) will be key for measuring TRGB distances out to further distances, where crowding will become an increasingly larger factor. Additionally, $K-$corrections become relevant for TRGB distances at the $1\%$ level for galaxies $\gtrsim 50$\,Mpc distant \citep{Anderson2022}.

We also present a preliminary measurement of the TRGB in NGC~1559 and NGC~5584, as a showcase of JWST's capabilities. While the repeatability of the singular HST TRGB measurement in NGC~5584 is a matter of some debate \citep{2015ApJ...807..133J, 2019ApJ...882...34F, 2022ApJ...932...15A}, the JWST TRGB measurement is clearly visible, even to the unassisted eye. In the near future, we will use our absolute TRGB calibration to measure well-tested distances to four supernova hosts from GO–1685 (NGC~1448, NGC~1559, NGC~5584, NGC~5643), which are host to a total of six type Ia supernovae. Combined with anticipated TRGB measurements from the upcoming Cycle 2 program GO-2875 \citep{2023jwst.prop.2875R}, an independent appraisal of the Cepheid distance scale \citep{2022ApJ...934L...7R} is well within reach with just the data already on JWST's schedule. 

Our calibration will also be useful as a preliminary absolute calibration of the TRGB + Surface Brightness Fluctuation (SBF) distance scale, where the SBF measurements would take the place of Type Ia supernovae in the final rung of the distance ladder \citep{2023arXiv230703116C}. In-progress observations from the Cycle 2 JWST Program GO-3055 \citep{2023jwst.prop.3055T} will provide the data needed to determine a secure TRGB calibration for the SBF distance scale. It is incredible to imagine that except for the initial geometric scaling, all other steps of the distance ladder can be obtained with NIRCam, including SBF measurements out into the Hubble flow. This would have the effect of nullifying many mutual sources of error along our proposed Population II distance ladder. 

To re-iterate, we do not present this as a final calibration of the TRGB in NIRCam's F090W filter. There will be inevitable updates to photometric reference files, zeropoints, and flat-fields, as well as changes to best practices for PSF photometry with DOLPHOT. Updates to the TRGB calibration presented here from our team will duly follow. What we do present is a well-tested set of measurements in NGC~4258 which form the basis of an initial calibration, which we believe will serve the community well in performing general distance measurements with the TRGB and JWST. Additionally, we are aware of at least one other program (GO-1638; \citealt{2021jwst.prop.1638M}) which will present not only a calibration in F090W, but also in a broader range of JWST filters. It is indeed an exciting time for work on the extragalactic distance scale.

\begin{acknowledgments}

We are indebted to all of those who spent years and even decades bringing {\it JWST} to fruition. 

GSA acknowledges financial support from JWST GO-1685. RLB is supported by the National Science Foundation through grant number AST-2108616. DIM, LNM and IDK are supported by the grant \textnumero~075--15--2022--262 (13.MNPMU.21.0003) of the Ministry of Science and Higher Education of the Russian Federation. RIA is funded by the SNSF via an Eccellenza Professorial Fellowship PCEFP2\_194638 and acknowledges support from the European Research Council (ERC) under the European Union's Horizon 2020 research and innovation programme (Grant Agreement No. 947660).

This research made use of the NASA Astrophysics Data System. This work is based on observations made with the NASA/ESA/CSA JWST. The data were obtained from the Mikulski Archive for Space Telescopes at the Space Telescope Science Institute, which is operated by the Association of Universities for Research in Astronomy, Inc., under NASA contract NAS 5-03127 for JWST. These observations are associated with program $\#$1685. The JWST data presented in this article were obtained from the Mikulski Archive for Space Telescopes (MAST) at the Space Telescope Science Institute. The specific observations analyzed can be accessed via \dataset[DOI: 10.17909/9ha5-p277]{https://doi.org/10.17909/9ha5-p277}. 

The Digitized Sky Surveys were produced at the Space Telescope Science Institute under U.S. Government grant NAG W-2166. The images of these surveys are based on photographic data obtained using the Oschin Schmidt Telescope on Palomar Mountain and the UK Schmidt Telescope. The plates were processed into the present compressed digital form with the permission of these institutions.

\end{acknowledgments}

\facility{JWST (NIRCam)}

\clearpage

\bibliography{paper}{}
\bibliographystyle{aasjournal}



\end{document}